\providecommand{\reserveinserts}[1]{}
\documentclass[AMA,Times1COL]{WileyNJDv5}

\articletype{Research Article}
\received{}
\revised{}
\accepted{}
\jname{Risk Analysis}
\journal{Risk Analysis}

\usepackage{amsmath,amssymb,amsthm,mathtools}
\usepackage{bm}
\usepackage{graphicx}
\usepackage{booktabs}
\usepackage{longtable}
\usepackage{array}
\usepackage{adjustbox}
\usepackage{multirow}
\usepackage{xcolor}
\usepackage{tikz}
\usetikzlibrary{arrows.meta,positioning,fit,calc,decorations.pathreplacing,backgrounds}
\usepackage{tcolorbox}
\tcbuselibrary{breakable}
\usepackage{enumitem}
\usepackage{listings}
\usepackage{setspace}

\numberwithin{equation}{section}

\makeatletter
\@ifundefined{theorem}{}{}
\@ifundefined{proposition}{}{}
\@ifundefined{remark}{}{}
\makeatother

\newtcolorbox{algorithmbox}[1]{
 colback=gray!5,
 colframe=black!65,
 title=#1,
 fonttitle=\bfseries,
 breakable,
 sharp corners,
 boxrule=0.6pt,
 left=7pt,
 right=7pt,
 top=7pt,
 bottom=7pt,
 before upper={\small\setlength{\parindent}{0pt}\setlength{\parskip}{3pt}}
}

\title{Probability Bound Analysis for Dependence Uncertainty in Risk and Decision Models}

\author[1,2]{Rowan Iskandar}
\address[1]{Medtronic International Trading Sarl, Tolochenaz, Vaud, Switzerland}
\address[2]{Center for Evidence Synthesis in Health, Brown University School of Public Health, Providence, RI, USA}
\corres{Rowan Iskandar, Rue des Amandiers 1, 2000 Neuchatel, Switzerland. Email: rowan.iskandar@gmail.com}

\abstract{Risk and decision models often combine sparse marginal information, precisely specified probability distributions, and dependence assumptions that are only partly justified. Probability bound analysis (PBA) represents epistemic uncertainty through probability boxes, but many applications assume independence or require dependence structures to be fully specified. We develop a dependence-sensitive PBA framework for black-box risk and decision models in which both marginal information and dependence information may be incomplete. The framework combines p-box parameters, precise-CDF parameters, and fixed quantities; incorporates specified dependence through copulas; and propagates unknown dependence through Fr\'echet-style admissible coupling sets. We also extend the construction to cross-dependence between imprecisely specified and precisely specified inputs. In an illustrative risk decision model, dependence assumptions materially affected output bounds and tail-risk summaries; analyses that ignored or simplified dependence produced narrower characterizations of plausible outcomes. The framework supports transparent uncertainty propagation when evidence is insufficient to justify either precise marginal distributions or a single dependence model.}
\keywords{probability bound analysis; p-box; dependence uncertainty; Fr\'echet bounds; risk analysis}

\begin{document}
\doublespacing
\maketitle

\section{Introduction}
Risk and decision models are routinely used to support choices under incomplete information \citep{ohagan2006uncertain,troffaes2012robust,augustin2014introduction}. In domains such as public health, environmental management, engineering safety, biosecurity, and technology assessment, analysts often know that a model input lies within a plausible range or satisfies a small set of moment constraints, but they may not be able to justify a fully specified probability distribution \citep{walley1991statistical,ferson2009pbox}. Dependence among inputs is even harder to justify \citep{ferson2004,nelsen2006copulas,joe2014copulas}. Parameters may share data sources, biological or physical mechanisms, expert judgments, or structural constraints, yet the available evidence may not identify a single joint distribution.

This combination of imprecise marginal information and uncertain dependence creates a central problem for risk analysis. Standard probabilistic sensitivity analysis propagates a single joint probability model through the system \citep{oakley2004,claxton2005psa,jackson2019voi}. When the marginal distributions or dependence assumptions are weakly supported, however, a single joint model can create a misleading impression of precision \citep{walley1991statistical,augustin2014introduction}. Conversely, ignoring dependence by imposing independence can assign mass to implausible input combinations and can understate or distort the range of plausible risk outcomes \citep{ferson2004,jacques2006,mara2012,neine2020}. A risk analysis framework should therefore distinguish what is known about each marginal input from what is known, partially known, or unknown about their dependence \citep{ferson2004,ruschendorf1991frechet,troffaes2014lower}.

Probability bound analysis (PBA) is one such approach.
It is an imprecise probability model for representing and propagating uncertainty when precise probability distributions for model parameters cannot be justified \citep{walley1991statistical,ferson2009pbox,ferson2004,iskandar2021}. 
Instead of requiring a single (\textit{precise}) cumulative distribution function (CDF), PBA represents uncertainty with a pair of bounding CDFs, often called a probability box (p-box) \citep{ferson2009pbox}. 
This representation is useful when available information consists only of sparse summary information such as minima, maxima, means, medians, or standard deviations \citep{ferson2004,ferson2009pbox,iskandar2021pba}.
In many of its formulations, PBA assumes independence among model parameters \citep{iskandar2021,ferson2004}. 
It is a very common assumption and, unfortunately, rarely receives scrutiny.
Dependence is common in risk and decision models \citep{oakley2004,jacques2006,mara2012}. 
For example, treatment benefits and adverse-event risks may be biologically linked, and parameters estimated from the same study may be statistically correlated \citep{baio2011,neine2020}. 
In these settings, the independence assumption can create parameter combinations that are implausible and may lead to underestimation of uncertainty bounds \citep{ferson2004,mara2012}.
The methodological gap is therefore twofold. 
Existing PBA methods can represent imprecise marginal information, but dependence has often been handled by assuming independence. 
Conversely, copulas and related dependence models are widely used with precise marginal distributions, but they do not by themselves allow imprecision in the marginal information \citep{nelsen2006copulas,joe2014copulas}. 
A model may require both components: uncertain marginals and uncertain or partially specified dependence \citep{walley1991statistical,ruschendorf1991frechet,troffaes2014lower}.
This study develops a generalized dependence-sensitive PBA framework for risk and decision models that addresses this methodological gap. 
First, we briefly review the concept of PBA and its uncertainty propagation.
Second, we formally describe two approaches to modeling dependence between parameters whose uncertainty is represented by probability boxes: using a copula and Fr\'echet-style bounds for unknown dependence.
Then, we extend the two approaches to model cross-dependence between different types of parameters.
Fourth, we demonstrate how to implement the PBA algorithm using a worked risk-model example.
Lastly, we conclude with a discussion on the limitations and directions for future research.
Throughout this exposition, we try to strike a balance between mathematical rigor and accessibility to practitioners.

\section{Preliminaries}

We consider a model as a black-box mapping
\begin{equation}
Y=\mathcal{M}(\boldsymbol{\theta}),
\label{eq:blackbox_model}
\end{equation}
where \(\boldsymbol{\theta}\in\Theta\) denotes the vector of model parameters, \(\Theta\) is the joint parameter space, and \(Y\) denotes a scalar decision-relevant model outcome, such as cost, net monetary benefit, or survival. 
The black-box formulation is intentionally general: we need only evaluate \(\mathcal{M}\) at specified parameter values, and our framework does not require a closed-form expression for the distribution of \(Y\).
We partition the parameter vector as
\begin{equation}
\boldsymbol{\theta}
=
(\boldsymbol{\theta}_b,\boldsymbol{\eta},\boldsymbol{\phi}),
\label{eq:parameter_partition}
\end{equation}
where \(\boldsymbol{\theta}_b=(\theta_1,\ldots,\theta_K)\) denotes $K$ parameters represented using p-boxes (\textit{p-box parameters}), \(\boldsymbol{\eta}=(\eta_1,\ldots,\eta_M)\) denotes $M$ parameters represented using precise cumulative distribution functions (\textit{precise-CDF parameters}), and \(\boldsymbol{\phi}\) denotes fixed parameters. Precise-CDF parameters are propagated by sampling values from their precisely-specified distributions. 
Fixed parameters do not require uncertainty propagation, although their values still affect the model output. 
P-box parameters require a different construction because their marginal distributions are not specified precisely.
The goal of parameter uncertainty quantification is to represent uncertainty in these inputs and propagate it through \(\mathcal{M}\) to characterize the induced uncertainty in \(Y\). 
The following subsections define p-boxes and demonstrate how p-box parameters are propagated through $\mathcal{M}$, which in turns provide a way to incorporate a dependence structure.

\subsection{Probability boxes}

For each p-box parameter \(\theta_i \in \boldsymbol{\theta}_b\), uncertainty is represented by
\begin{equation}
\mathcal{P}_{\mathcal{D}_i}(\theta_i)
=
[\underline{F}_i(\theta_i\mid\mathcal{D}_i),\overline{F}_i(\theta_i\mid\mathcal{D}_i)],
\label{eq:pbox_definition}
\end{equation}
where \(\underline{F}_i\) and \(\overline{F}_i\) are lower and upper bounds on the unknown cumulative distribution function of \(\theta_i\), and \(\mathcal{D}_i\) denotes the available information about that parameter. 
Under the PBA framework, for every realization of $\theta$, we can only assign an \textit{interval of CDF values},  $\left[\underline{F}_i,\overline{F}_i\right]$, in contrast to one CDF value.
In sparse-data settings, \(\mathcal{D}_i\) may include only support bounds, such as a minimum and maximum, or it may additionally include summary information such as a mean, median, standard deviation, or selected quantiles. Each subset of \(\mathcal{D}_i\) induces a different p-box: with only support information the bounding functions are wide, whereas additional information tightens the region between \(\underline{F}_i\) and \(\overline{F}_i\). 
In the limiting case where the CDF is fully specified, the two bounding functions coincide and the p-box degenerates to a precise CDF. 
Formally, the bounds defining a p-box have the following properties:
\begin{enumerate}
    \item $\underline{F}_i$ and $\overline{F}_i$ are CDFs
    \item \(\underline{F}_i(\theta_i\mid\mathcal D_i) \le F_i(\theta_i) \le \overline{F}_i(\theta_i\mid\mathcal D_i)\) for all \(\theta_i\) in the support
    \item $\underline{F}_i$ and $\overline{F}_i$ form the sharpest bounds implied by \(\mathcal D_i\)
    \item \(\underline{F}_i\) and \(\overline{F}_i\) are consistent with \(\mathcal D_i\)
\end{enumerate}
Here, a CDF of \(\theta_i\) is consistent with the minimal data \(\mathcal D_i\) if each element in \(\mathcal D_i\) can be equated to a statistic derivable from that CDF.
supplementary material summarizes several commonly used free p-box constructions from minimal data.

\subsection{Discretization, marginal slice masses, and hyperrectangles}
\label{sec:discretization}

\begin{figure}[!htbp]
\centering
\resizebox{\textwidth}{!}{%
\begin{tikzpicture}[
 font=\normalsize,
 stepbox/.style={draw, rounded corners, fill=gray!10, align=center, minimum height=0.95cm, inner sep=5pt},
 arrow/.style={-{Latex[length=2.2mm]}, thick},
 massarrow/.style={-{Latex[length=2.2mm]}, thick, dashed},
 note/.style={align=center, font=\small}
]


\node[stepbox, minimum width=5.3cm] at (0,5.15)
{\textbf{Step 1: discretize marginal p-boxes into slices}\\[1mm]
Probability scale for each parameter};

\draw[->] (-2.3,2.95) -- (2.3,2.95) node[right] {\(u\)};
\foreach \x/\lab in {-2.3/0,-1.15/{c_i^j},0/{d_i^j},1.15/{},2.3/1} {
 \draw (\x,2.87) -- (\x,3.03);
 \node[below] at (\x,2.82) {\(\lab\)};
}
\draw[fill=blue!15, draw=blue!70, thick] (-1.15,2.65) rectangle (0,3.25);
\node[above] at (-0.575,3.30) {\([c_i^j,d_i^j]\)};
\node[note] at (0,1.55) {\normalsize Slice mass\\\normalsize \(m_i^j=d_i^j-c_i^j\)};

\node[stepbox, minimum width=5.3cm] at (6.4,5.15)
{\textbf{Step 2: combine slices}\\[1mm]
Probability-space hyperrectangle};

\draw[->] (4.3,1.55) -- (8.5,1.55) node[right] {\(U_1\)};
\draw[->] (4.3,1.55) -- (4.3,4.05) node[above] {\(U_2\)};
\foreach \x in {4.3,5.35,6.4,7.45,8.5} {
 \draw[gray!45] (\x,1.55) -- (\x,4.05);
}
\foreach \y in {1.55,2.175,2.8,3.425,4.05} {
 \draw[gray!45] (4.3,\y) -- (8.5,\y);
}
\draw[fill=green!18, draw=green!60!black, very thick] (5.35,2.175) rectangle (6.4,2.8);
\node[note,font=\large] at (7.05,2.49) {\(\mathcal U_{\mathbf{k}}\)};
\node[note,font=\large] at (6.4,0.95) {\normalsize Mass assignment is defined\\ \normalsize over the \(\mathcal U_{\mathbf{k}}\)'s};

\node[stepbox, minimum width=5.3cm] at (12.8,5.15)
{\textbf{Step 3: map to parameter space}\\[1mm]
Parameter-space hyperrectangle};

\draw[->] (10.7,1.55) -- (14.9,1.55) node[right] {\(\theta_1\)};
\draw[->] (10.7,1.55) -- (10.7,4.05) node[above] {\(\theta_2\)};
\foreach \x in {10.7,11.75,12.8,13.85,14.9} {
 \draw[gray!45] (\x,1.55) -- (\x,4.05);
}
\foreach \y in {1.55,2.175,2.8,3.425,4.05} {
 \draw[gray!45] (10.7,\y) -- (14.9,\y);
}
\draw[fill=orange!20, draw=orange!80!black, very thick] (11.75,2.175) rectangle (12.8,2.8);
\node[note] at (13.45,2.49) {\(\mathcal H_{\mathbf{k}}\)};
\node[note] at (12.8,0.85) {\normalsize Model optimized\\ \normalsize over \(\mathcal H_{\mathbf{k}}\)};

\draw[arrow] (2.65,3.15) -- (4.0,3.15)
 node[midway, above, note] {\normalsize Cartesian\\\normalsize product};
\draw[arrow] (8.85,3.15) -- (10.4,3.15)
 node[midway, above, note] {\normalsize quasi-inverse\\\normalsize mapping};


\node[stepbox, minimum width=15.6cm] at (6.4,-0.35)
{\textbf{Step 4: choose a dependence assumption and assign masses}\\[1mm]
The \(\mathcal U_{\mathbf{k}}\)'s are fixed by marginal discretization; our dependence assumption determines the masses \(p_{\mathbf{k}}\).};

\node[note] at (6.4,-1.45)
{\large Fixed probability-space cells};

\foreach \x/\name/\mass/\col in {2.8/{\(\mathcal U_1\)}/{\(p_1\)}/blue!12,
 5.2/{\(\mathcal U_2\)}/{\(p_2\)}/green!16,
 7.6/{\(\mathcal U_3\)}/{\(p_3\)}/yellow!20,
 10.0/{\(\mathcal U_4\)}/{\(p_4\)}/red!12} {
 \draw[fill=\col, draw=black, thick] (\x,-2.85) rectangle +(1.35,0.72);
 \node at (\x+0.675,-2.47) {\name};
 \node[below] at (\x+0.675,-2.85) {\mass};
}

\node[note,font=\large] at (6.4,-3.55)
{\normalsize Masses \(p_{\mathbf{k}}\) are computed differently depending on the dependence assumption};

\draw[massarrow] (6.4,-3.05) -- (6.4,-3.35);

\node[stepbox, minimum width=4.2cm, fill=blue!8] (indep) at (0.8,-5.25)
{\textbf{Independence}\\[1mm]
\(p_{\mathbf{k}}=\prod_{i} m_i^{k_i}\)};

\node[stepbox, minimum width=4.2cm, fill=green!8] (copula) at (6.4,-5.25)
{\textbf{Specified copula}\\[1mm]
\(p_{\mathbf{k}}=P_C(\mathcal U_{\mathbf{k}})\)};

\node[stepbox, minimum width=4.2cm, fill=orange!10] (unknown) at (13.0,-5.25)
{\textbf{Unknown dependence}\\[1mm]
\(p_{\mathbf{k}}\) optimized over feasible masses};

\draw[massarrow] (6.4,-3.75) -- (indep.north);
\draw[massarrow] (6.4,-3.75) -- (copula.north);
\draw[massarrow] (6.4,-3.75) -- (unknown.north);

\node[note, align=center] at (6.4,-6.57)
{\large All three approaches attach masses to the same \(\mathcal U_{\mathbf{k}}\)'s;\\
\large the corresponding \(\mathcal H_{\mathbf{k}}\)'s are then evaluated through the model.};

\end{tikzpicture}%
}
\caption{Construction of probability-space hyperrectangles \(\mathcal U_{\mathbf{k}}\), parameter-space hyperrectangles \(\mathcal H_{\mathbf{k}}\), and dependence-sensitive mass assignment. Marginal p-box discretization first creates probability slices and interval-valued parameter realizations. The Cartesian products of probability slices define \(\mathcal U_{\mathbf{k}}\), while the corresponding Cartesian products of parameter intervals define \(\mathcal H_{\mathbf{k}}\). Dependence assumptions determine how the masses \(p_{\mathbf{k}}\) are assigned to the fixed probability-space cells \(\mathcal U_{\mathbf{k}}\). These masses may be computed by independence, by a specified copula, or by optimization over admissible masses under unknown dependence; the assumptions do not change the marginal intervals or the geometry of the \(\mathcal H_{\mathbf{k}}\)'s.}
\label{fig:hyperrectangle_construction}
\end{figure}
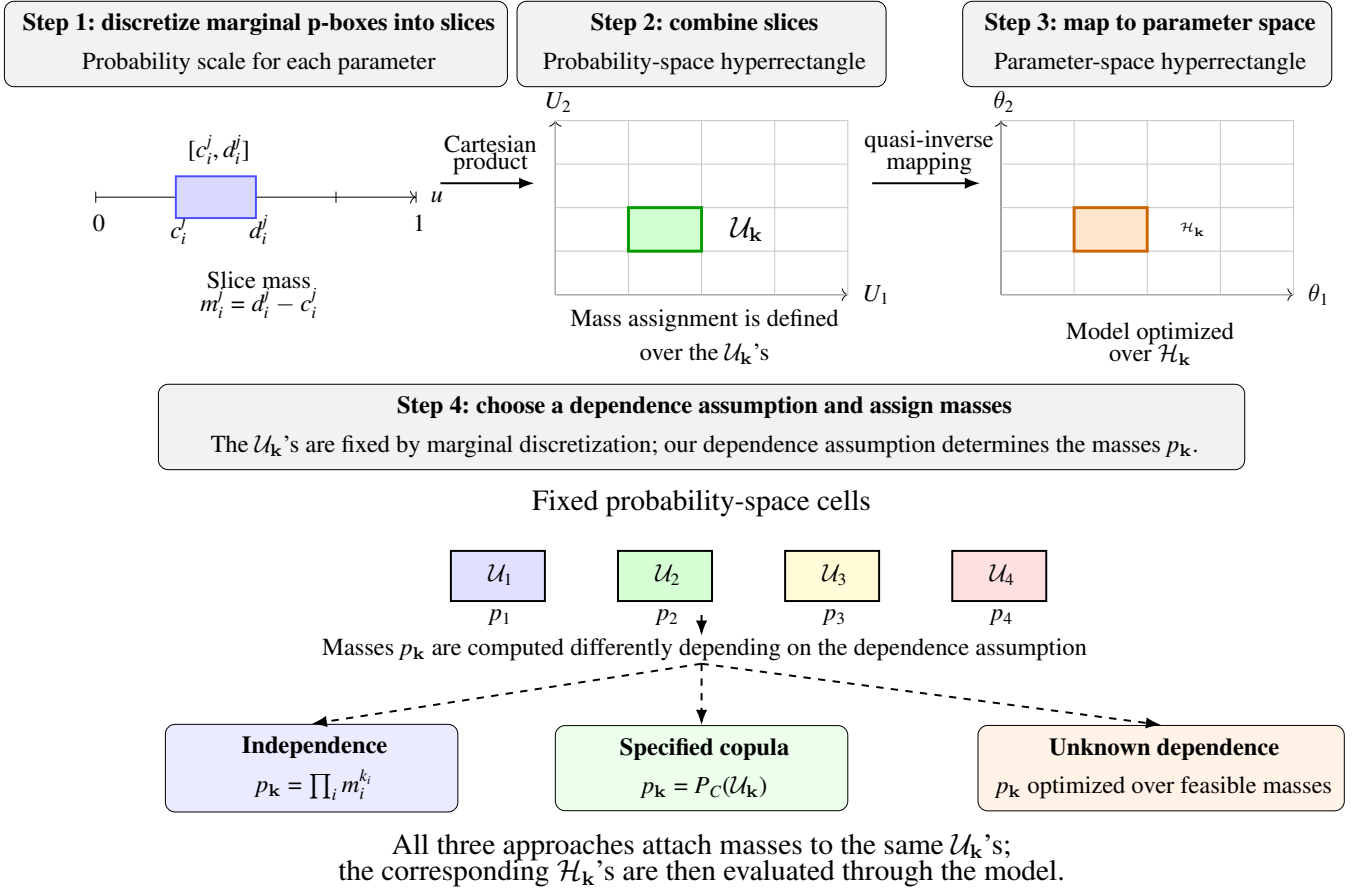

A p-box represents a set of cumulative distribution functions rather than a single distribution. Thus, propagating a p-box through a black-box model is not as simple as drawing one value from a specified distribution (as in PSA). 
In principle, we must characterize the range of model outputs that could arise from all distributions consistent with the p-box  on the model parameters.
Hence, sampling a p-box is akin to sampling an interval.
This is achieved using a finite-dimensional approximation of the probability domain (see Iskandar \citep{iskandar2021}).
Discretization partitions the probability scale of each p-box parameter into slices, maps each probability slice into an interval of plausible parameter values, and then evaluates the model over the resulting regions of the input space. 
As the number of slices increases, the accuracy of the discretized representation improves, though at the cost of higher computational burden.
For each p-box parameter \(\theta_i\in\boldsymbol{\theta}_b\), 
we partition the probability interval \([0,1]\) into \(n_i\) slices,
\begin{equation}
[c_i^1,d_i^1],
[c_i^2,d_i^2],
\ldots,
[c_i^{n_i},d_i^{n_i}],
\label{eq:probability_partition}
\end{equation}
where the \(j\)-th slice has probability-scale width
\begin{equation}
m_i^j=d_i^j-c_i^j.
\label{eq:marginal_slice_mass}
\end{equation}
The quantity \(m_i^j\) is the marginal mass assigned to the \(j\)-th probability slice of parameter \(i\). 
Each probability slice is then mapped from probability space to parameter space using the quasi-inverses of the p-box:
\begin{equation}
a_i^j=\overline{F}_i^{-1}(c_i^j\mid\mathcal{D}_i),
\qquad
b_i^j=\underline{F}_i^{-1}(d_i^j\mid\mathcal{D}_i).
\label{eq:outer_discretization}
\end{equation}
Here, \(F^{-1}(p)\) denotes the generalized inverse \(\inf\{x:F(x)\ge p\}\) because the bounding functions may have flat portions or jumps after discretization.
This pair of mappings produce the interval-valued realization \([a_i^j,b_i^j]\) for parameter \(\theta_i\). 
The interval contains the parameter values that may correspond to the probability slice \([c_i^j,d_i^j]\) under any cumulative distribution function lying within the p-box.
To propagate several p-box parameters jointly, let $\mathbf{k}=(k_1,\ldots,k_K)$ index one selected slice for each of the \(K\) p-box parameters. 
The Cartesian product of the corresponding parameter intervals defines the parameter-space hyperrectangle
\begin{equation}
\mathcal{H}_{\mathbf{k}}
=
\prod_{i=1}^{K}[a_i^{k_i},b_i^{k_i}],
\label{eq:parameter_hyperrectangle}
\end{equation}
and the corresponding product of probability slices defines the probability-space hyperrectangle
\begin{equation}
\mathcal{U}_{\mathbf{k}}
=
\prod_{i=1}^{K}[c_i^{k_i},d_i^{k_i}].
\label{eq:probability_hyperrectangle}
\end{equation}
Define the following multi-index set:
\[
\mathcal K
=
\left\{
(k_1,\ldots,k_K):
1\le k_i\le n_i
\text{ for } i=1,\ldots,K
\right\},
\]
where each index vector \(\mathbf{k}\in\mathcal K\) identifies one combination of marginal probability slices across the \(K\) p-box parameters. Thus, \(\mathcal K\) indexes both the probability-space hyperrectangles \(\mathcal U_{\mathbf{k}}\) and the corresponding parameter-space hyperrectangles \(\mathcal H_{\mathbf{k}}\).
In sum, the (marginal) discretization defines the slice masses \(m_i^j\), the interval-valued realizations \([a_i^j,b_i^j]\), the probability-space hyperrectangles \(\mathcal{U}_{\mathbf{k}}\), and the corresponding parameter-space hyperrectangles \(\mathcal{H}_{\mathbf{k}}\). 
The set of parameter-space hyperrectangles \(\{\mathcal{H}_{\mathbf{k}}\}\) forms the computational domain over which the p-box parameters are propagated through the black-box model \(\mathcal{M}\). 
We denote by \(p_{\mathbf{k}}\) the probability mass assigned to probability-space hyperrectangle \(\mathcal U_{\mathbf{k}}\), equivalently, the mass associated with the corresponding parameter-space hyperrectangle \(\mathcal H_{\mathbf{k}}\). 
Dependence assumptions enter only after these marginal discretization components have been constructed. 
Specifically, dependence assumptions affect \(p_{\mathbf{k}}\), that is, how probability mass is assigned across the probability-space hyperrectangles \(\mathcal U_{\mathbf{k}}\); they do not alter the marginal slice masses \(m_i^j\) or the marginal intervals \([a_i^j,b_i^j]\). supplementary material provides concise mathematical justifications for the marginal-mass preservation and the validity of the constructed output p-box.
Figure~\ref{fig:hyperrectangle_construction} summarizes this construction and separates the geometric construction of \(\mathcal{H}_{\mathbf{k}}\) from the dependence-driven mass assignment over \(\mathcal{U}_{\mathbf{k}}\).

\section{Dependence Structures in PBA}
\label{sec:dependence_choice}

The marginal p-box construction in Section~\ref{sec:discretization} fixes the probability-scale slices, the corresponding parameter-space hyperrectangles, and the marginal slice masses. Dependence assumptions enter after this construction. They determine how probability mass is assigned across the fixed probability-space cells; they do not change the marginal p-boxes or the geometry of the hyperrectangles. The same construction is used throughout this section: first, specify or bound the cell masses; second, optimize the black-box model over the relevant parameter-space cells; and third, aggregate the cellwise output bounds into an output p-box. This separation is central for risk analysis because it distinguishes uncertainty about marginal evidence from uncertainty about dependence. We present an algorithm to conduct PBA under dependence uncertainty in supplementary material.

\subsection{Independence}

The marginal discretization described in Section~\ref{sec:discretization} produces three objects that remain fixed throughout the dependence analysis: the probability-space hyperrectangles \(\mathcal U_{\mathbf{k}}\), the corresponding parameter-space hyperrectangles \(\mathcal H_{\mathbf{k}}\), and the marginal slice masses \(m_i^j\). To complete the uncertainty propagation, probability mass must be assigned to the probability-space hyperrectangles. The simplest assignment assumes that the uncertain parameters are mutually independent.

Under independence, the joint mass assigned to \(\mathcal U_{\mathbf{k}}\) is the product of the selected marginal slice masses:
\begin{equation}
p_{\mathbf{k}}^{\perp}
=
\prod_{i=1}^{K}
m_i^{k_i}
=
\prod_{i=1}^{K}
\left(
d_i^{k_i}
-
c_i^{k_i}
\right).
\label{eq:independence_mass}
\end{equation}

Equation~\eqref{eq:independence_mass} uniquely determines the joint probability mass associated with every probability-space hyperrectangle. Consequently, no additional dependence model or mass-assignment optimization is required. The uncertainty propagation proceeds by evaluating the black-box model over \(\mathcal H_{\mathbf{k}}\) and aggregating the resulting outputs using \(p_{\mathbf{k}}^{\perp}\). The remaining sections relax the independence assumption in progressively more general ways. Section~\ref{sec:copula} considers specified dependence among p-box parameters using a copula. Section~\ref{sec:frechet} considers unknown dependence among p-box parameters by optimizing over admissible couplings. Section~\ref{sec:cross_dependence} extends both ideas to cross-dependence between p-box parameters and precise-CDF parameters. Independence is therefore a special case of the broader dependence framework.

\subsection{Copula-based dependence}
\label{sec:copula}

Copulas provide a flexible way to model dependence while preserving the marginal p-box construction. A copula separates marginal behavior from dependence. For \(K\) random variables with marginal cumulative distribution functions \(F_1,\ldots,F_K\), Sklar's theorem gives
\begin{equation}
F(\theta_1,\ldots,\theta_K)
=
C\{F_1(\theta_1),\ldots,F_K(\theta_K)\},
\label{eq:sklar_copula}
\end{equation}
where \(C:[0,1]^K\rightarrow[0,1]\) is a copula. The copula describes how the parameters co-vary after their marginal distributions have been transformed onto the common probability scale \([0,1]\). In PBA, the marginal distributions are not fixed because each p-box represents a set of admissible cumulative distribution functions. After discretization, however, the probability-space cells \(\mathcal U_{\mathbf{k}}\) and the corresponding parameter-space hyperrectangles \(\mathcal H_{\mathbf{k}}\) are fixed. Dependence assumptions therefore affect the masses assigned to these cells rather than the cell geometry itself. The copula approach follows the same PBA sequence as independence: the cells are fixed, the dependence model assigns their masses, and the model is then propagated over the associated parameter-space hyperrectangles.

Let \(U_i\in[0,1]\) denote the probability-scale coordinate associated with p-box parameter \(\theta_i\). A copula specifies
\begin{equation}
C(u_1,\ldots,u_K)
=
\Pr(U_1\le u_1,\ldots,U_K\le u_K).
\label{eq:copula_distribution}
\end{equation}
The copula gives cumulative probability up to a corner of the unit hypercube. PBA, however, requires probability mass inside a probability-space hyperrectangle. For the probability-space hyperrectangle \(\mathcal U_{\mathbf{k}}\) defined in Equation~\eqref{eq:probability_hyperrectangle}, define, for each corner vector \(\mathbf{s}=(s_1,\ldots,s_K)\in\{0,1\}^K\),
\begin{equation}
u_i^{s_i}
=
\begin{cases}
c_i^{k_i}, & s_i=0,\\
d_i^{k_i}, & s_i=1.
\end{cases}
\label{eq:copula_corner_coordinate}
\end{equation}
The copula-induced mass of \(\mathcal U_{\mathbf{k}}\) is
\begin{equation}
w_{\mathbf{k}}^C
=
\Pr(\mathcal U_{\mathbf{k}})
=
\sum_{\mathbf{s}\in\{0,1\}^K}
(-1)^{K-\sum_{i=1}^{K} s_i}
C(u_1^{s_1},\ldots,u_K^{s_K}).
\label{eq:copula_rectangle_mass}
\end{equation}
This is the inclusion--exclusion increment of the copula over the probability-space cell. In the specified-copula case, \(w_{\mathbf{k}}^C\) is the realized value of the general cell mass \(p_{\mathbf{k}}\). For the independence copula
\begin{equation}
C_\perp(u_1,\ldots,u_K)=\prod_{i=1}^{K}u_i,
\label{eq:independence_copula}
\end{equation}
Equation~\eqref{eq:copula_rectangle_mass} reduces to Equation~\eqref{eq:independence_mass}.

After the cell masses have been assigned, the model is propagated through the parameter-space hyperrectangles. Let
\begin{equation}
\boldsymbol{\eta}^{(1)},\ldots,\boldsymbol{\eta}^{(N)}
\label{eq:precise_cdf_draws}
\end{equation}
be Monte Carlo draws from the joint distribution of the precise-CDF parameters. For each draw \(\boldsymbol{\eta}^{(l)}\), \(l=1,\ldots,N\), and each p-box hyperrectangle \(\mathcal H_{\mathbf{k}}\), the black-box model is minimized and maximized over the p-box parameters while holding the precise-CDF draw and fixed parameters fixed:
\begin{equation}
\underline Y_{\mathbf{k},l}
=
\min_{\boldsymbol{\theta}_b\in\mathcal H_{\mathbf{k}}}
\mathcal M(\boldsymbol{\theta}_b,\boldsymbol{\eta}^{(l)},\boldsymbol{\phi}),
\qquad
\overline Y_{\mathbf{k},l}
=
\max_{\boldsymbol{\theta}_b\in\mathcal H_{\mathbf{k}}}
\mathcal M(\boldsymbol{\theta}_b,\boldsymbol{\eta}^{(l)},\boldsymbol{\phi}).
\label{eq:copula_hyperrectangle_optimization_precise_cdf}
\end{equation}
If no precise-CDF parameters are present, the outer Monte Carlo averaging step is omitted. Given the copula-induced masses \(w_{\mathbf{k}}^C\), the output p-box is obtained by averaging over the outer Monte Carlo draws and summing over the p-box hyperrectangles:
\begin{equation}
\underline F_Y(y)
=
\frac{1}{N}
\sum_{l=1}^{N}
\sum_{\mathbf{k}\in\mathcal K}
w_{\mathbf{k}}^C
\mathbf 1\{\overline Y_{\mathbf{k},l}\le y\},
\label{eq:copula_output_pbox_lower_precise_cdf}
\end{equation}
and
\begin{equation}
\overline F_Y(y)
=
\frac{1}{N}
\sum_{l=1}^{N}
\sum_{\mathbf{k}\in\mathcal K}
w_{\mathbf{k}}^C
\mathbf 1\{\underline Y_{\mathbf{k},l}\le y\}.
\label{eq:copula_output_pbox_upper_precise_cdf}
\end{equation}

\subsection{Fr\'echet-style bounds for unknown dependence}
\label{sec:frechet}

Copulas require a specified dependence model. When no defensible dependence model is available, the p-box cell masses cannot be computed from a copula or from the independence assumption. Instead, the masses \(p_{\mathbf{k}}\) introduced in Section~\ref{sec:discretization} are treated as unknown. The marginal discretization nevertheless remains fixed: the probability-space cells \(\mathcal U_{\mathbf{k}}\) and the marginal slice masses \(m_i^j\) are already determined. Consequently, although the joint masses \(p_{\mathbf{k}}\) are unknown, they cannot be arbitrary. Any admissible mass assignment must preserve the marginal slice masses implied by the discretization. Specifically, the total mass assigned to all cells whose \(i\)-th coordinate corresponds to slice \(j\) must equal \(m_i^j\). Together with non-negativity and unit total mass, these marginal-preservation constraints define the admissible coupling set
\begin{equation}
\mathcal Q
=
\left\{
p=(p_{\mathbf{k}})_{\mathbf{k}\in\mathcal K}:
p_{\mathbf{k}}\ge0,
\sum_{\mathbf{k}\in\mathcal K}p_{\mathbf{k}}=1,
\sum_{\mathbf{k}:k_i=j}p_{\mathbf{k}}=m_i^j
\text{ for all } i,j
\right\}.
\label{eq:frechet_coupling_set}
\end{equation}
Once the admissible mass set has been specified, the remaining tasks parallel Section~\ref{sec:copula}, except that the final mass-assignment step is optimized over \(\mathcal Q\) rather than computed from a specified copula.

Before optimizing over \(\mathcal Q\), the model must first be propagated over each fixed hyperrectangle. This first optimization layer is
\begin{equation}
\underline Y_{\mathbf{k},l}
=
\min_{\boldsymbol{\theta}_b\in\mathcal H_{\mathbf{k}}}
\mathcal M(\boldsymbol{\theta}_b,\boldsymbol{\eta}^{(l)},\boldsymbol{\phi}),
\qquad
\overline Y_{\mathbf{k},l}
=
\max_{\boldsymbol{\theta}_b\in\mathcal H_{\mathbf{k}}}
\mathcal M(\boldsymbol{\theta}_b,\boldsymbol{\eta}^{(l)},\boldsymbol{\phi}).
\label{eq:frechet_hyperrectangle_optimization_precise_cdf}
\end{equation}
For a fixed \(p\in\mathcal Q\) and threshold \(y\in\mathbb R\), define
\begin{equation}
\underline F_Y(y;p)
=
\frac{1}{N}
\sum_{l=1}^{N}
\sum_{\mathbf{k}\in\mathcal K}
p_{\mathbf{k}}
\mathbf 1\{\overline Y_{\mathbf{k},l}\le y\},
\label{eq:frechet_candidate_lower_output}
\end{equation}
and
\begin{equation}
\overline F_Y(y;p)
=
\frac{1}{N}
\sum_{l=1}^{N}
\sum_{\mathbf{k}\in\mathcal K}
p_{\mathbf{k}}
\mathbf 1\{\underline Y_{\mathbf{k},l}\le y\}.
\label{eq:frechet_candidate_upper_output}
\end{equation}
The indicator terms in Equations~\eqref{eq:frechet_candidate_lower_output} and \eqref{eq:frechet_candidate_upper_output} are fixed after Equation~\eqref{eq:frechet_hyperrectangle_optimization_precise_cdf} has been computed. The remaining uncertainty is only the allocation of probability mass across the fixed cells. The Fr\'echet bounds are therefore
\begin{equation}
\underline F_Y(y)
=
\min_{p\in\mathcal Q}
\underline F_Y(y;p),
\label{eq:frechet_lower_output_bound}
\end{equation}
and
\begin{equation}
\overline F_Y(y)
=
\max_{p\in\mathcal Q}
\overline F_Y(y;p).
\label{eq:frechet_upper_output_bound}
\end{equation}
Equations~\eqref{eq:frechet_lower_output_bound} and \eqref{eq:frechet_upper_output_bound} are finite-dimensional linear programs because the objective functions and all constraints are linear in \(p_{\mathbf{k}}\). supplementary material gives the corresponding validity and sharpness arguments.

\subsection{Cross-dependence between p-box and precise-CDF parameters}
\label{sec:cross_dependence}

Sections~\ref{sec:copula} and~\ref{sec:frechet} address dependence among p-box parameters only. In many risk models, however, dependence may also exist between parameters represented by p-boxes and parameters represented by precise cumulative distribution functions. This creates a different problem. P-box parameters are propagated through discretized hyperrectangles \(\mathcal H_{\mathbf{k}}\), whereas precise-CDF parameters are often propagated through Monte Carlo draws from fully specified distributions. Dependence between these two classes of uncertainty therefore cannot be represented using the p-box hyperrectangles alone.

\begin{figure}[t]
\centering
\resizebox{\textwidth}{!}{%
\begin{tikzpicture}[
 font=\normalsize,
 block/.style={draw, rounded corners, fill=gray!10, align=center, minimum height=0.95cm, inner sep=8pt},
 inputblock/.style={draw, rounded corners, fill=white, align=center, minimum height=0.85cm, inner sep=5pt},
 arrow/.style={-{Latex[length=2.4mm]}, thick},
 note/.style={align=center, font=\small}
]

\node[block, minimum width=17.0cm] (title) at (7.0,7.15)
{\textbf{General cross-dependence structure}\\
P-box parameters \(\boldsymbol{\theta}_b=(\theta_1,\ldots,\theta_K)\) and precise-CDF parameters \(\boldsymbol{\eta}=(\eta_1,\ldots,\eta_M)\) may be cross-dependent.};

\node[block, minimum width=5.8cm, fill=blue!6] (pbox_head) at (1.8,5.45)
{\textbf{P-box parameters}\\
\(\boldsymbol{\theta}_b=(\theta_1,\ldots,\theta_K)\)};

\node[block, minimum width=5.8cm, fill=green!8] (precise_head) at (12.2,5.45)
{\textbf{Precise-CDF parameters}\\
\(\boldsymbol{\eta}=(\eta_1,\ldots,\eta_M)\)};

\node[inputblock, minimum width=5.8cm] (pbox_rep) at (1.8,3.85)
{Discretize p-boxes into\\
\(\mathcal U_{\mathbf{k}}\) and \(\mathcal H_{\mathbf{k}}\)};

\node[inputblock, minimum width=5.8cm] (precise_rep) at (12.2,3.85)
{Use precise CDFs \(F_{\eta_h}\)\\
to sample or discretize \(\eta_h\)};

\node[block, minimum width=8.2cm, fill=yellow!12] (probscale) at (7.0,0.85)
{\textbf{Common probability scale}\\
P-box coordinates \(U_1,\ldots,U_K\) and precise-CDF coordinates \(V_1,\ldots,V_M\)};

\node[block, minimum width=8.2cm, fill=purple!10] (cross) at (7.0,-1.05)
{\textbf{Cross-dependence model}\\
Specifies or bounds how \((U_1,\ldots,U_K)\) varies with \((V_1,\ldots,V_M)\)};

\node[block, minimum width=6.0cm, fill=orange!12] (copula) at (2.8,-3.75)
{\textbf{Specified cross-dependence}\\
Conditional copula masses \(w_{\mathbf{k}\mid l}^{C}\)};

\node[block, minimum width=6.0cm, fill=cyan!12] (frechet) at (11.2,-3.75)
{\textbf{Unknown cross-dependence}\\
Joint Fr\'echet mass optimization};

\node[note, align=center] at (7.0,-5.35)
{The marginal p-box and precise-CDF information are kept fixed.\\
Cross-dependence determines how joint probability mass is arranged across combined parameter cells.};

\draw[arrow] (pbox_head.south) -- (pbox_rep.north);
\draw[arrow] (precise_head.south) -- (precise_rep.north);

\coordinate (prob_in_left) at ([xshift=-3.0cm]probscale.north);
\coordinate (prob_in_right) at ([xshift=3.0cm]probscale.north);
\draw[arrow] (pbox_rep.south) -- ++(0,-0.65) -| (prob_in_left);
\draw[arrow] (precise_rep.south) -- ++(0,-0.65) -| (prob_in_right);

\draw[arrow] (probscale.south) -- (cross.north);

\coordinate (cross_out_left) at ([xshift=-2.4cm]cross.south);
\coordinate (cross_out_right) at ([xshift=2.4cm]cross.south);
\draw[arrow] (cross_out_left) -- ++(0,-0.55) -| (copula.north);
\draw[arrow] (cross_out_right) -- ++(0,-0.55) -| (frechet.north);

\end{tikzpicture}%
}
\caption{General structure of cross-dependence between p-box parameters and precise-CDF parameters. P-box parameters are represented by probability-space cells \(\mathcal U_{\mathbf{k}}\) and parameter-space hyperrectangles \(\mathcal H_{\mathbf{k}}\). Precise-CDF parameters are represented through their cumulative distribution functions \(F_{\eta_h}\), either by outer sampling or by discretization. Cross-dependence is modeled on the common probability scale using \(U_i\) for p-box coordinates and \(V_h\) for precise-CDF coordinates. The copula approach specifies conditional masses \(w_{\mathbf{k}\mid l}^{C}\), whereas the joint Fr\'echet approach optimizes over admissible joint mass assignments.}
\label{fig:cross_dependence_overview}
\end{figure}
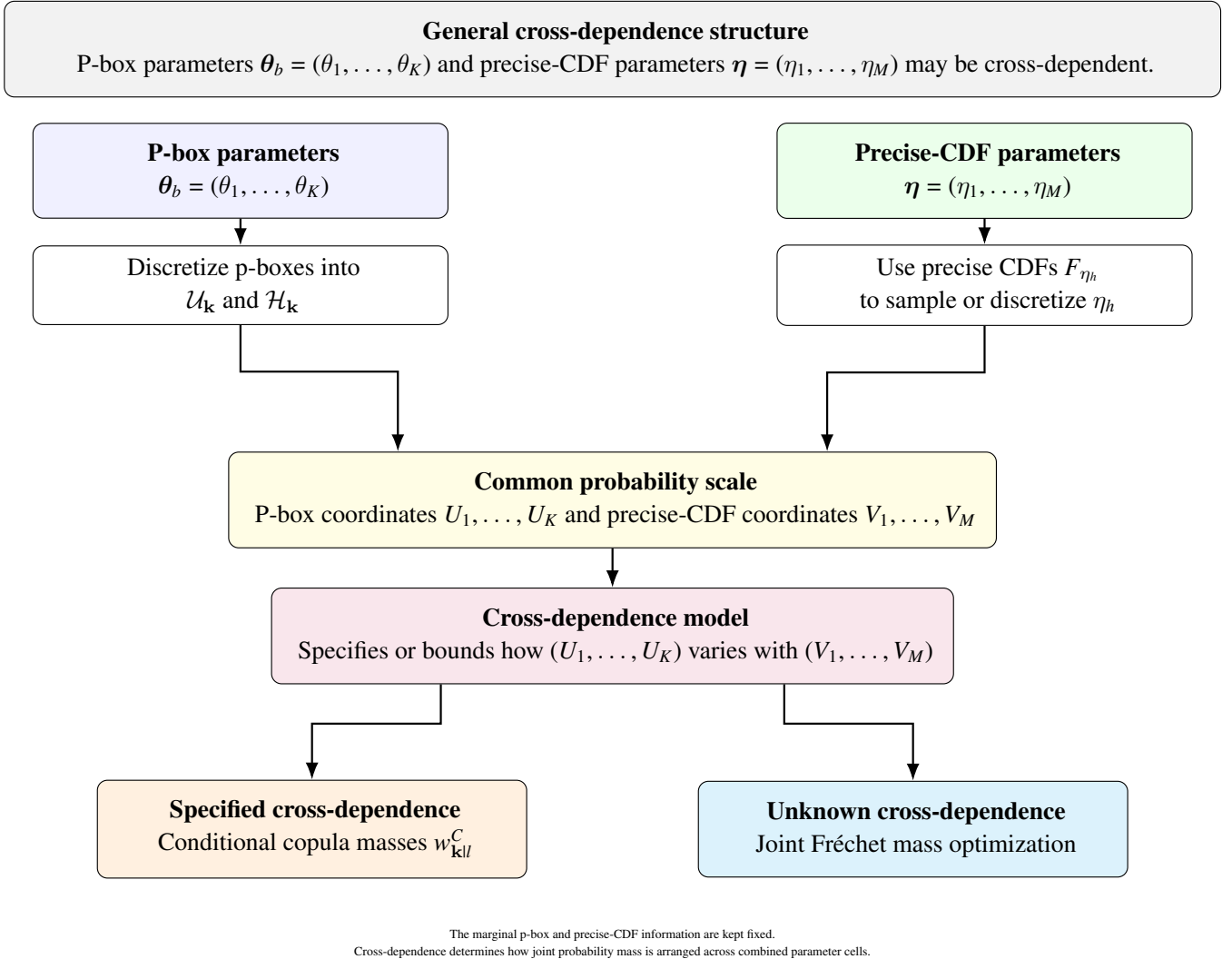

A common probability-scale representation is therefore required. P-box parameters use coordinates \(U_i\in[0,1]\). Precise-CDF parameters use
\begin{equation}
V_h=F_{\eta_h}(\eta_h),
\qquad h=1,\ldots,M.
\label{eq:precise_probability_scale_coordinate}
\end{equation}
The resulting variables \(V_h\) are uniform on \([0,1]\). Cross-dependence can then be specified or bounded for the combined vector \((U_1,\ldots,U_K,V_1,\ldots,V_M)\). Sections~\ref{sec:cross_copula} and~\ref{sec:cross_frechet} extend the specified-dependence and unknown-dependence constructions to this mixed representation. Figure~\ref{fig:cross_dependence_overview} summarizes this construction.

\subsubsection{Specified cross-dependence using a copula}
\label{sec:cross_copula}

A specified cross-dependence model can be represented by a joint copula \(C_{bc}\) for \((U_1,\ldots,U_K,V_1,\ldots,V_M)\). This is the cross-dependence analogue of Section~\ref{sec:copula}: the dependence model is specified, so the relevant masses are calculated rather than optimized. For each outer draw \(\boldsymbol{\eta}^{(l)}\), define its probability-scale value
\begin{equation}
\boldsymbol v^{(l)}
=
\left(
F_{\eta_1}(\eta_1^{(l)}),
\ldots,
F_{\eta_M}(\eta_M^{(l)})
\right).
\label{eq:precise_outer_draw_probability_scale}
\end{equation}
Conditional on this value, the mass assigned to p-box probability-space cell \(\mathcal U_{\mathbf{k}}\) is
\begin{equation}
w_{\mathbf{k}\mid l}^C
=
\Pr\left\{
(U_1,\ldots,U_K)\in\mathcal U_{\mathbf{k}}
\,\middle|\,
(V_1,\ldots,V_M)=\boldsymbol v^{(l)}
\right\}.
\label{eq:conditional_copula_mass}
\end{equation}
Thus, unlike Section~\ref{sec:copula}, the p-box cell weights may vary across outer Monte Carlo draws; \(w_{\mathbf{k}\mid l}^C\) is the draw-specific mass assigned to \(\mathcal U_{\mathbf{k}}\) conditional on \(\boldsymbol v^{(l)}\). If the joint copula has density \(c_{bc}(\boldsymbol u,\boldsymbol v)\), this conditional mass can be expressed as
\begin{equation}
w_{\mathbf{k}\mid l}^C
=
\frac{
\int_{\mathcal U_{\mathbf{k}}}
c_{bc}(\boldsymbol u,\boldsymbol v^{(l)})
\,d\boldsymbol u
}{
c_c(\boldsymbol v^{(l)})
},
\label{eq:conditional_copula_density_mass}
\end{equation}
where \(c_c(\boldsymbol v^{(l)})\) is the marginal copula density of the precise-CDF probability-scale vector \((V_1,\ldots,V_M)\). This expression is evaluated only at values for which \(c_c(\boldsymbol v^{(l)})>0\). Conditioning on \(\boldsymbol v^{(l)}\) redistributes probability mass across the fixed p-box cells. The model is then optimized over the same p-box-derived hyperrectangles as before, with \(\underline Y_{\mathbf{k},l}\) and \(\overline Y_{\mathbf{k},l}\) defined as in Equation~\eqref{eq:copula_hyperrectangle_optimization_precise_cdf}; only the weights attached to those hyperrectangles depend on the precise-CDF draw. For \(y\in\mathbb R\), the output p-box is constructed by replacing unconditional copula weights \(w_{\mathbf{k}}^C\) with conditional weights \(w_{\mathbf{k}\mid l}^C\):
\begin{equation}
\underline F_Y(y)
=
\frac{1}{N}
\sum_{l=1}^{N}
\sum_{\mathbf{k}\in\mathcal K}
w_{\mathbf{k}\mid l}^C
\mathbf 1\{\overline Y_{\mathbf{k},l}\le y\},
\label{eq:cross_copula_output_lower}
\end{equation}
and
\begin{equation}
\overline F_Y(y)
=
\frac{1}{N}
\sum_{l=1}^{N}
\sum_{\mathbf{k}\in\mathcal K}
w_{\mathbf{k}\mid l}^C
\mathbf 1\{\underline Y_{\mathbf{k},l}\le y\}.
\label{eq:cross_copula_output_upper}
\end{equation}
Figure~\ref{fig:cross_dependence_copula} illustrates this conditional mass-assignment mechanism.

\begin{figure}[!htbp]
\centering
\resizebox{0.86\textwidth}{!}{%
\begin{tikzpicture}[
  font=\scriptsize,
  titlebox/.style={draw, rounded corners, fill=gray!10, align=center, minimum width=12.8cm, minimum height=0.95cm, inner sep=8pt},
  stepbox/.style={draw, rounded corners, align=center, minimum width=8.9cm, minimum height=0.98cm, inner sep=8pt},
  sidebox/.style={draw, rounded corners, align=center, minimum width=4.9cm, minimum height=0.95cm, inner sep=8pt},
  arrow/.style={-{Latex[length=2.7mm,width=1.7mm]}, very thick, shorten >=3pt, shorten <=3pt},
  note/.style={align=center, font=\scriptsize}
]


\node[stepbox, fill=blue!2] (draw) at (0,8.1)
{\textbf{1. Draw precise-CDF parameters}\\
\(\boldsymbol{\eta}^{(l)}=(\eta_1^{(l)},\ldots,\eta_M^{(l)})\)};

\node[stepbox, fill=blue!7] (prob) at (0,6.0)
{\textbf{2. Transform the draw to probability scale}\\
\(\boldsymbol v^{(l)}=(F_{\eta_1}(\eta_1^{(l)}),\ldots,F_{\eta_M}(\eta_M^{(l)}))\)};

\node[stepbox, fill=purple!9] (copula) at (0,3.9)
{\textbf{3. Condition the joint copula}\\
\(C_{bc}\) is evaluated conditional on \(\boldsymbol V=\boldsymbol v^{(l)}\)};

\node[stepbox, fill=green!9] (mass) at (0,1.75)
{\textbf{4. Assign conditional masses to fixed p-box cells}\\
\(\mathcal U_{\mathbf{k}} \mapsto w_{\mathbf{k}\mid l}^{C}\)};

\node[note] (celllabel) at (0,0.00)
{Fixed probability-space cells with draw-specific weights};

\foreach \x/\name/\wt/\col in {-4.5/{\(\mathcal U_1\)}/{\(w_{1\mid l}^{C}\)}/green!10,
                              -1.5/{\(\mathcal U_2\)}/{\(w_{2\mid l}^{C}\)}/green!16,
                               1.5/{\(\mathcal U_3\)}/{\(w_{3\mid l}^{C}\)}/green!22,
                               4.5/{\(\mathcal U_4\)}/{\(w_{4\mid l}^{C}\)}/green!28} {
  \draw[fill=\col, draw=black, thick] (\x-0.75,-1.30) rectangle +(1.5,0.78);
  \node at (\x,-0.89) {\name};
  \node[below] at (\x,-1.30) {\scriptsize \wt};
}

\node[sidebox, fill=orange!10] (eval) at (-3.8,-3.75)
{\textbf{5a. Optimize on \(\mathcal H_{\mathbf{k}}\)}\\
compute \(\underline Y_{\mathbf{k},l}\), \(\overline Y_{\mathbf{k},l}\)};

\node[sidebox, fill=orange!10] (agg) at (3.8,-3.75)
{\textbf{5b. Aggregate outputs}\\
use \(w_{\mathbf{k}\mid l}^{C}\) to form\\
\(\underline F_Y(y)\), \(\overline F_Y(y)\)};

\node[note] at (0,-5.45)
{The copula changes the weights attached to the cells, not the geometry of \(\mathcal U_{\mathbf{k}}\) or \(\mathcal H_{\mathbf{k}}\).};

\draw[arrow] (draw.south) -- (prob.north);
\draw[arrow] (prob.south) -- (copula.north);
\draw[arrow] (copula.south) -- (mass.north);
\draw[arrow] (mass.south) -- ($(celllabel.north)+(0,0.10)$);
\coordinate (split) at (0,-2.20);
\draw[arrow] (0,-1.45) -- (split);
\draw[arrow] (split) -| (eval.north);
\draw[arrow] (split) -| (agg.north);

\draw[arrow] (eval.east) -- (agg.west);

\end{tikzpicture}%
}
\caption{Copula-based cross-dependence between p-box parameters and precise-CDF parameters. For each outer draw \(\boldsymbol{\eta}^{(l)}\), the precise-CDF parameters are mapped to probability scale as \(\boldsymbol v^{(l)}\). Conditioning the joint copula \(C_{bc}\) on \(\boldsymbol V=\boldsymbol v^{(l)}\) gives conditional masses \(w_{\mathbf{k}\mid l}^{C}\) for the fixed p-box probability-space cells \(\mathcal U_{\mathbf{k}}\). These masses are then used to aggregate the optimized model-output bounds \(\underline Y_{\mathbf{k},l}\) and \(\overline Y_{\mathbf{k},l}\) over the corresponding parameter-space hyperrectangles \(\mathcal H_{\mathbf{k}}\).}
\label{fig:cross_dependence_copula}
\end{figure}

\subsubsection{Unknown cross-dependence using a Fr\'echet-style formulation}
\label{sec:cross_frechet}

When no defensible cross-copula is available, the precise-CDF parameters can be discretized and included in a joint Fr\'echet-style coupling. This is the cross-dependence analogue of Section~\ref{sec:frechet}: the cells are constructed first, the black-box model is optimized within each joint cell, and the admissible mass assignment is optimized after those cellwise bounds have been computed. For each precise-CDF parameter \(\eta_h\), partition the probability scale into \(R_h\) slices,
\begin{equation}
J_h^s=[e_h^s,f_h^s],
\qquad
\mu_h^s=f_h^s-e_h^s,
\qquad s=1,\ldots,R_h.
\label{eq:cross_precise_probability_slice}
\end{equation}
The corresponding parameter-space interval is
\begin{equation}
\mathcal G_h^s
=
\left[
F_{\eta_h}^{-1}(e_h^s),
F_{\eta_h}^{-1}(f_h^s)
\right].
\label{eq:cross_precise_quantile_cell}
\end{equation}
Let
\[
\mathcal R
=
\left\{
(r_1,\ldots,r_M):
1\le r_h\le R_h
\text{ for } h=1,\ldots,M
\right\}.
\]
For a vector of precise-CDF slice indices \(\mathbf{r}=(r_1,\ldots,r_M)\in\mathcal R\), parallel to the p-box multi-index \(\mathbf{k}\), define
\begin{equation}
\mathcal G_{\mathbf{r}}
=
\prod_{h=1}^{M}
\mathcal G_h^{r_h}.
\label{eq:cross_precise_cell_product}
\end{equation}
Thus, \(\mathcal G_{\mathbf{r}}\) is the precise-CDF analogue of the p-box hyperrectangle \(\mathcal H_{\mathbf{k}}\). The combined joint cell is
\begin{equation}
\mathcal J_{\mathbf{k},\mathbf{r}}
=
\mathcal H_{\mathbf{k}}
\times
\mathcal G_{\mathbf{r}}.
\label{eq:cross_joint_cell}
\end{equation}
The index pair \((\mathbf{k},\mathbf{r})\in\mathcal K\times\mathcal R\) identifies one cell in the joint discretization of all uncertain parameters.

The first optimization layer propagates each joint cell through the black-box model. For every \((\mathbf{k},\mathbf{r})\), compute
\begin{equation}
\underline Y_{\mathbf{k},\mathbf{r}}
=
\min_{\boldsymbol{\theta}_b\in\mathcal H_{\mathbf{k}},\;\boldsymbol{\eta}\in\mathcal G_{\mathbf{r}}}
\mathcal M(\boldsymbol{\theta}_b,\boldsymbol{\eta},\boldsymbol{\phi}),
\qquad
\overline Y_{\mathbf{k},\mathbf{r}}
=
\max_{\boldsymbol{\theta}_b\in\mathcal H_{\mathbf{k}},\;\boldsymbol{\eta}\in\mathcal G_{\mathbf{r}}}
\mathcal M(\boldsymbol{\theta}_b,\boldsymbol{\eta},\boldsymbol{\phi}).
\label{eq:cross_joint_cell_optimization}
\end{equation}
This step determines the lower and upper model outputs attainable inside each joint cell. It is separate from the dependence problem: no joint probability masses are optimized in Equation~\eqref{eq:cross_joint_cell_optimization}.

The second optimization layer assigns joint masses to the cells. Let \(p_{\mathbf{k},\mathbf{r}}\) denote the unknown mass assigned to \(\mathcal J_{\mathbf{k},\mathbf{r}}\). The admissible set is
\begin{equation}
\mathcal Q_{bc}
=
\left\{
p:
p_{\mathbf{k},\mathbf{r}}\ge0,
\sum_{\mathbf{k}\in\mathcal K}\sum_{\mathbf{r}\in\mathcal R}p_{\mathbf{k},\mathbf{r}}=1,
\sum_{\substack{\mathbf{k}\in\mathcal K\\ k_i=j}}\sum_{\mathbf{r}\in\mathcal R}p_{\mathbf{k},\mathbf{r}}=m_i^j
\text{ for all }i,j,
\sum_{\substack{\mathbf{r}\in\mathcal R\\ r_h=s}}\sum_{\mathbf{k}\in\mathcal K}p_{\mathbf{k},\mathbf{r}}=\mu_h^s
\text{ for all }h,s
\right\}.
\label{eq:cross_frechet_feasible_set}
\end{equation}
The first set of marginal constraints preserves the p-box slice masses. The second preserves the precise-CDF probability-slice masses. Thus, \(\mathcal Q_{bc}\) varies only the admissible dependence structure while preserving the marginal information.

\begin{figure}[!htbp]
\centering

\resizebox{\textwidth}{!}{%
\begin{tikzpicture}[
 font=\normalsize,
 block/.style={draw, rounded corners, fill=gray!10, align=center, minimum height=1.05cm, inner sep=7pt},
 arrow/.style={-{Latex[length=2.8mm]}, thick},
 dashedarrow/.style={-{Latex[length=2.8mm]}, thick, dashed},
 note/.style={align=center, font=\small}
]

\node[block, minimum width=4.2cm, fill=blue!6] (pbtitle) at (0,6.2)
{P-box discretization\\\(\mathcal H_{\mathbf{k}},\mathcal U_{\mathbf{k}}\)};
\draw[->] (-1.9,1.75) -- (2.0,1.75) node[right] {\(U_1\)};
\draw[->] (-1.9,1.75) -- (-1.9,3.85) node[above] {\(U_2\)};
\foreach \x in {-1.9,-0.95,0,0.95,1.9} {
 \draw[gray!45] (\x,1.75) -- (\x,3.85);
}
\foreach \y in {1.75,2.275,2.8,3.325,3.85} {
 \draw[gray!45] (-1.9,\y) -- (1.9,\y);
}
\draw[fill=blue!15, draw=blue!70, thick] (-0.95,2.275) rectangle (0,2.8);
\node[note] at (0.35,2.55) {\(\mathcal U_{\mathbf{k}}\)};

\node[block, minimum width=4.2cm, fill=green!8] (pctitle) at (12.8,6.2)
{Precise-CDF discretization\\\(\mathcal G_{\mathbf{r}}\)};
\draw[->] (10.9,2.8) -- (14.7,2.8) node[right] {\(V_h\)};
\foreach \x/\lab in {10.9/0,11.85/{e_h^s},12.8/{f_h^s},13.75/{},14.7/1} {
 \draw (\x,2.72) -- (\x,2.88);
 \node[below] at (\x,2.67) {\(\lab\)};
}
\draw[fill=green!18, draw=green!60!black, thick] (11.85,2.5) rectangle (12.8,3.1);
\node[note] at (12.35,3.25) {\(\mathcal G_h^s\)};
\node[note] at (12.8,1.65) {Products across \(h\) give \(\mathcal G_{\mathbf{r}}\)};

\node[block, minimum width=5.8cm, fill=orange!10] (joint) at (6.4,4.25)
{Joint cell\\
\(\mathcal J_{\mathbf{k},\mathbf{r}}=\mathcal H_{\mathbf{k}}\times\mathcal G_{\mathbf{r}}\)};
\node[block, minimum width=5.8cm, fill=orange!10] (mass) at (6.4,1.95)
{Unknown joint mass\\
\(p_{\mathbf{k},\mathbf{r}}\)};

\node[block, minimum width=12.2cm, fill=white] (Q) at (6.4,-1.15)
{\(\begin{aligned}
\mathcal Q_{bc}
=
\{(p_{\mathbf{k},\mathbf{r}}):\;&
p_{\mathbf{k},\mathbf{r}}\ge0,\;
\sum_{\mathbf{k}\in\mathcal K}\sum_{\mathbf{r}\in\mathcal R}p_{\mathbf{k},\mathbf{r}}=1,\\
&
\sum_{\substack{\mathbf{k}\in\mathcal K\\ k_i=j}}\sum_{\mathbf{r}\in\mathcal R} p_{\mathbf{k},\mathbf{r}}=m_i^j,\;
\sum_{\substack{\mathbf{r}\in\mathcal R\\ r_h=s}}\sum_{\mathbf{k}\in\mathcal K} p_{\mathbf{k},\mathbf{r}}=\mu_h^{s}
\}
\end{aligned}\)};

\draw[arrow] (pbtitle.east) -- ++(0.8,0) |- (joint.west);
\draw[arrow] (pctitle.west) -- ++(-0.8,0) |- (joint.east);
\draw[arrow] (joint.south) -- ++(0,-0.95) -- (mass.north);
\draw[arrow] (mass.south) -- ++(0,-0.95) -- (Q.north);

\path (0,-3.8);
\end{tikzpicture}%
}
\caption{Unknown cross-dependence using a joint Fr\'echet formulation. The precise-CDF parameters are discretized into probability slices and mapped to intervals \(\mathcal G_h^s\). Products of these intervals define \(\mathcal G_{\mathbf{r}}\), with \(\mathbf{r}\in\mathcal R\), which are combined with p-box-derived hyperrectangles \(\mathcal H_{\mathbf{k}}\), \(\mathbf{k}\in\mathcal K\), to form joint cells \(\mathcal J_{\mathbf{k},\mathbf{r}}\). The unknown masses \(p_{\mathbf{k},\mathbf{r}}\) are optimized over the feasible set \(\mathcal Q_{bc}\), which preserves the marginal slice masses for both p-box and precise-CDF parameters.}
\label{fig:cross_dependence_frechet}
\end{figure}
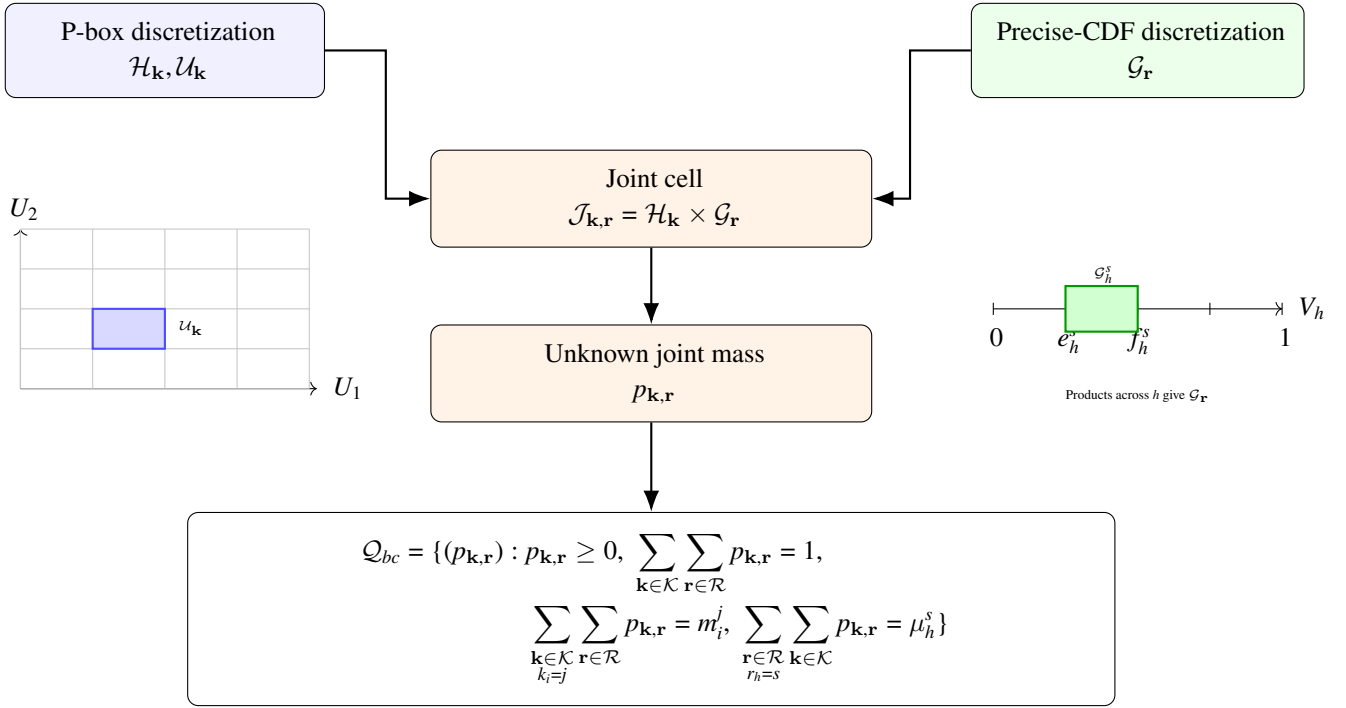

For a fixed \(p\in\mathcal Q_{bc}\) and threshold \(y\in\mathbb R\), the candidate output bounds are
\begin{equation}
\underline F_Y(y;p)
=
\sum_{\mathbf{k}\in\mathcal K}\sum_{\mathbf{r}\in\mathcal R}
p_{\mathbf{k},\mathbf{r}}
\mathbf 1\{\overline Y_{\mathbf{k},\mathbf{r}}\le y\},
\label{eq:cross_candidate_lower_output}
\end{equation}
and
\begin{equation}
\overline F_Y(y;p)
=
\sum_{\mathbf{k}\in\mathcal K}\sum_{\mathbf{r}\in\mathcal R}
p_{\mathbf{k},\mathbf{r}}
\mathbf 1\{\underline Y_{\mathbf{k},\mathbf{r}}\le y\}.
\label{eq:cross_candidate_upper_output}
\end{equation}
The indicator terms in Equations~\eqref{eq:cross_candidate_lower_output} and \eqref{eq:cross_candidate_upper_output} are fixed after Equation~\eqref{eq:cross_joint_cell_optimization} has been computed. The remaining unknowns are the masses \(p_{\mathbf{k},\mathbf{r}}\). The Fr\'echet lower and upper output bounds under unknown cross-dependence are obtained by solving
\begin{equation}
\underline F_Y(y)
=
\min_{p\in\mathcal Q_{bc}}
\underline F_Y(y;p),
\label{eq:cross_frechet_lower_output_bound}
\end{equation}
and
\begin{equation}
\overline F_Y(y)
=
\max_{p\in\mathcal Q_{bc}}
\overline F_Y(y;p).
\label{eq:cross_frechet_upper_output_bound}
\end{equation}
These are finite-dimensional linear programs because the objectives and all constraints are linear in the unknown masses \(p_{\mathbf{k},\mathbf{r}}\). Figure~\ref{fig:cross_dependence_frechet} summarizes this construction.

\section{Illustrative risk-model application}
\label{sec:illustrative_application_intro}
We adapt the animal herd inspection model of Troffaes and Gosling as an illustrative risk-decision model~\cite{troffaes2012robust} to demonstrate the introduced dependence-sensitive PBA approaches.
The decision problem concerns choosing the number of animals to test ($m$) before allowing the herd of imported animals ($n$) to pass inspection. 
The herd contains an unknown number \(d\) of diseased animals. 
The diagnostic test has sensitivity \(p\) and specificity \(q\). 
Testing \(m\) animals incurs cost \(c(m)\). 
If diseased animals pass inspection undetected, an outbreak cost \(a(d)\) is incurred. 
If at least one diseased animal is detected, the herd is terminated at cost \(t(n)\). 
Details on the computational model are presented in supplementary material. 
The uncertain quantities are grouped according to the three parameter classes in the framework.
The p-box parameters are the infection probability \(r\) and the outbreak cost \(a\). 
The precise-CDF parameters are sensitivity \(p\) and specificity \(q\). 
Sensitivity and specificity are treated as dependent parameters as they are characteristics of the same diagnostic test. 
The fixed parameters are the herd size \(n\), the coefficients in \(c(m)\), the termination cost multiplier, and the loss threshold used for exceedance summaries.
We consider eleven scenarios. 
For each scenario we calculate the expected loss and the probability of the loss exceeding a threshold.
Scenarios 1--9 vary the dependence assumptions within the two parameter pairs \((r,a)\) and \((p,q)\). 
Scenario 10 introduces Fr\'echet unknown cross-dependence between the pairs \((r,a)\) and \((p,q)\). 
Scenario 11 introduces a Gaussian-copula cross-dependence between the same pair.
As a comparison, we conduct PSA assigning precise CDF to all uncertain parameters. 
\subsection{Scenario 10 worked implementation}
Scenario~10 applies the generalized algorithm from supplementary material with independence for \((r,a)\), Gaussian-copula dependence for \((p,q)\) with \(\rho_{pq}=-0.85\), and Fr\'echet-type unknown cross-dependence between the pair blocks \((r,a)\) and \((p,q)\). The implementation is easiest to follow by keeping four objects separate. First, construct the two within-block cell systems: \(400\) cells for \((r,a)\) and \(400\) cells for \((p,q)\). Second, compute the within-block mass vectors \(w^{ra}=(w^{ra}_u)\) and \(w^{pq}=(w^{pq}_v)\). Third, optimize the \(400\times400\) cross-block coupling matrix \(\Pi^\times=(\Pi^\times_{uv})\). Fourth, use the precomputed cell-output bounds \(\underline Y_{uv}\) and \(\overline Y_{uv}\) in the threshold-specific linear programs. Thus, \(u\) indexes a flattened \((r,a)\) block cell, \(v\) indexes a flattened \((p,q)\) block cell, and \(\Pi^\times_{uv}\) is the unknown joint mass assigned to their combination.
\\ \\
\textbf{Step 1: Construct the marginal p-boxes.} The p-box for \(r\) was constructed using the range of prior means explored in the article, from \(0.0002\) to \(0.0032\) in addition to an assumed median value to avoid vacuous probability: $\mathcal D_r=\{r_L=0.0002,\;r_M=0.0016,\;r_U=0.0032\}$.
The available information for \(a\) has wide lower and upper bounds representing deep uncertainty in the cost of an outbreak: $\mathcal D_a=\{a_L=5{,}000{,}000,\;\mu_a=10{,}000{,}000,\;a_U=20{,}000{,}000\}$.
Using the p-box formulas for \(\mathcal D_r\), the p-boxes for $r$ are 
\[
\underline F_r(x)=
\begin{cases}
0, & x<0.0016,\\
0.5, & 0.0016\le x<0.0032,\\
1, & x\ge0.0032,
\end{cases}
\qquad
\overline F_r(x)=
\begin{cases}
0, & x<0.0002,\\
0.5, & 0.0002\le x<0.0016,\\
1, & x\ge0.0016.
\end{cases}
\]
Using the p-box formulas for \(\mathcal D_a\),
\[
\underline F_a(x)
=
\begin{cases}
0, & x<10{,}000{,}000,\\
\dfrac{x-10{,}000{,}000}{x-5{,}000{,}000}, & 10{,}000{,}000\le x<20{,}000{,}000,\\
1, & x\ge20{,}000{,}000,
\end{cases}
\]
and
\[
\overline F_a(x)
=
\begin{cases}
0, & x<5{,}000{,}000,\\
\dfrac{10{,}000{,}000}{20{,}000{,}000-x}, & 5{,}000{,}000\le x<10{,}000{,}000,\\
1, & x\ge10{,}000{,}000.
\end{cases}
\]
To ensure the comparison with PBA using the same support and median information, \(r\) is assigned a scaled beta distribution on \([0.0002,0.0032]\) with median \(0.0016\), which approximately gives: $\alpha_r=9.355,\qquad \beta_r=10.645$.
For the PSA comparator, the variable \(a\) is assigned a scaled beta distribution on \([5{,}000{,}000,20{,}000{,}000]\) with mean \(10{,}000{,}000\) and standard deviation equal to 20\% of the mean. 
The diagnostic parameters \(p\) and \(q\) remain precise-CDF parameters with beta distributions centered at 
\[
E(p)=0.85,\qquad E(q)=0.90,
\]
which give:
\[
p\sim\mathrm{Beta}(10.2,1.8),\qquad q\sim\mathrm{Beta}(10.8,1.2),
\]
\textbf{Step 2: Discretize the probability and parameter spaces.} For each parameter, $[0,1]$ is discretized into \(20\) probability slices:
\[
[c_i^j,d_i^j]
=
\left[
\frac{j-1}{20},\frac{j}{20}
\right],
\qquad
j=1,\ldots,20,
\]
so every marginal slice mass is
\[
m_i^j=d_i^j-c_i^j=0.05.
\]
For p-box parameters, each probability slice was mapped into the corresponding parameter-space interval using the quasi-inverse of p-box formulas
\[
[a_i^j,b_i^j]
=
\left[
\overline F_i^{-1}(c_i^j),
\underline F_i^{-1}(d_i^j)
\right].
\]
The quasi-inverses are
\[
\overline F_r^{-1}(u)=
\begin{cases}
0.0002, & 0\le u\le0.5,\\
0.0016, & 0.5<u\le1,
\end{cases}
\quad
\underline F_r^{-1}(u)=
\begin{cases}
0.0002, & u=0,\\
0.0016, & 0<u\le0.5,\\
0.0032, & 0.5<u\le1.
\end{cases}
\]
and
\[
\overline F_a^{-1}(u)
=
\begin{cases}
5{,}000{,}000, & 0\le u\le0.6667,\\
20{,}000{,}000-\dfrac{10{,}000{,}000}{u}, & 0.6667<u\le1.
\end{cases}
\]
The lower quasi-inverse used for the right endpoint is
\[
\underline F_a^{-1}(u)
=
\begin{cases}
5{,}000{,}000, & u=0,\\
\dfrac{10{,}000{,}000-5{,}000{,}000u}{1-u}, & 0<u<0.6667,\\
20{,}000{,}000, & 0.6667\le u\le1.
\end{cases}
\]
For the precise-CDF parameters, quantile-discretization intervals are used:
\[
[p_j^-,p_j^+]
=
\left[
F_p^{-1}\!\left(\frac{j-1}{20}\right),
F_p^{-1}\!\left(\frac{j}{20}\right)
\right],
\]
and
\[
[q_j^-,q_j^+]
=
\left[
F_q^{-1}\!\left(\frac{j-1}{20}\right),
F_q^{-1}\!\left(\frac{j}{20}\right)
\right].
\]
For example,
\[
j=1:\quad p\in[0.000,0.659],\qquad q\in[0.000,0.734],
\]
and
\[
j=20:\quad p\in[0.974,1.000],\qquad q\in[0.991,1.000].
\]
Table~\ref{tab:scenario10_slice_intervals} reports the slice intervals used in the stress-test implementation. These intervals are used only to compute finite Gaussian-copula cell masses and the cross-dependence. 
They do not represent imprecision in \(p\) or \(q\); the marginal distributions of \(p\) and \(q\) remain the precise beta CDFs throughout the PSA and PBA comparison.
\begin{table}[!htbp]
\centering
\caption{Scenario 10 slice intervals used for the stress-test implementation. The \(a\)-intervals are expressed in millions. The \(r\)- and \(a\)-columns are p-box-induced intervals, whereas the \(p\)- and \(q\)-columns are precise-CDF beta quantile intervals used for finite dependence calculations.}
\label{tab:scenario10_slice_intervals}
\scriptsize
\setlength{\tabcolsep}{3pt}
\begin{adjustbox}{max width=\textwidth}
\begin{tabular}{rccccc}
\toprule
Slice \(j\) & Probability interval & \(r\)-interval & \(a\)-interval & \(p\)-interval & \(q\)-interval\\
\midrule
1 & [0.00,0.05] & [0.0002,0.0016] & [5.000,10.263] & [0.000,0.659] & [0.000,0.734]\\
2 & [0.05,0.10] & [0.0002,0.0016] & [5.000,10.556] & [0.659,0.712] & [0.734,0.785]\\
3 & [0.10,0.15] & [0.0002,0.0016] & [5.000,10.882] & [0.712,0.747] & [0.785,0.816]\\
4 & [0.15,0.20] & [0.0002,0.0016] & [5.000,11.250] & [0.747,0.773] & [0.816,0.840]\\
5 & [0.20,0.25] & [0.0002,0.0016] & [5.000,11.667] & [0.773,0.794] & [0.840,0.859]\\
6 & [0.25,0.30] & [0.0002,0.0016] & [5.000,12.143] & [0.794,0.812] & [0.859,0.875]\\
7 & [0.30,0.35] & [0.0002,0.0016] & [5.000,12.692] & [0.812,0.829] & [0.875,0.888]\\
8 & [0.35,0.40] & [0.0002,0.0016] & [5.000,13.333] & [0.829,0.843] & [0.888,0.901]\\
9 & [0.40,0.45] & [0.0002,0.0016] & [5.000,14.091] & [0.843,0.857] & [0.901,0.912]\\
10 & [0.45,0.50] & [0.0002,0.0016] & [5.000,15.000] & [0.857,0.870] & [0.912,0.922]\\
11 & [0.50,0.55] & [0.0002,0.0032] & [5.000,16.111] & [0.870,0.882] & [0.922,0.931]\\
12 & [0.55,0.60] & [0.0016,0.0032] & [5.000,17.500] & [0.882,0.893] & [0.931,0.940]\\
13 & [0.60,0.65] & [0.0016,0.0032] & [5.000,19.286] & [0.893,0.904] & [0.940,0.948]\\
14 & [0.65,0.70] & [0.0016,0.0032] & [5.000,20.000] & [0.904,0.915] & [0.948,0.956]\\
15 & [0.70,0.75] & [0.0016,0.0032] & [5.714,20.000] & [0.915,0.926] & [0.956,0.963]\\
16 & [0.75,0.80] & [0.0016,0.0032] & [6.667,20.000] & [0.926,0.937] & [0.963,0.970]\\
17 & [0.80,0.85] & [0.0016,0.0032] & [7.500,20.000] & [0.937,0.948] & [0.970,0.977]\\
18 & [0.85,0.90] & [0.0016,0.0032] & [8.235,20.000] & [0.948,0.960] & [0.977,0.984]\\
19 & [0.90,0.95] & [0.0016,0.0032] & [8.889,20.000] & [0.960,0.974] & [0.984,0.991]\\
20 & [0.95,1.00] & [0.0016,0.0032] & [9.474,20.000] & [0.974,1.000] & [0.991,1.000]\\
\bottomrule
\end{tabular}
\end{adjustbox}
\end{table}
\\ \\
\textbf{Step 3, 4: Construct the finite parameter cells}
The full parameter cells combine p-box intervals for \((r,a)\) with beta-quantile intervals for \((p,q)\). To avoid confusing this implementation index with the p-box-only multi-index in Section~\ref{sec:discretization}, write the full four-dimensional implementation index as \(\boldsymbol{\kappa}=(k_r,k_a,k_p,k_q)\). The corresponding full parameter-space cell is
\[
\mathcal H^{\times}_{\boldsymbol{\kappa}}
=
[a_r^{k_r},b_r^{k_r}]
\times
[a_a^{k_a},b_a^{k_a}]
\times
[p_{k_p}^{-},p_{k_p}^{+}]
\times
[q_{k_q}^{-},q_{k_q}^{+}],
\]
where
\[
\boldsymbol{\kappa}=(k_r,k_a,k_p,k_q),
\]
and \(k_r,k_a,k_p,k_q\in\{1,\ldots,20\}\) are the selected slices for \(r,a,p,\) and \(q\), respectively.
The corresponding probability-scale rectangles are
\[
\mathcal U_{\boldsymbol{\kappa}}
=
[c_r^{k_r},d_r^{k_r}]
\times
[c_a^{k_a},d_a^{k_a}]
\times
[c_p^{k_p},d_p^{k_p}]
\times
[c_q^{k_q},d_q^{k_q}].
\]
The Cartesian product contains \(20\times20\times20\times20=160{,}000\) finite cells.
For instance, \(\boldsymbol{\kappa}=(4,12,16,9)\) means the fourth probability slice for \(r\), the twelfth probability slice for \(a\), the sixteenth quantile slice for \(p\), and the ninth quantile slice for \(q\):
\[
\mathcal H^{\times}_{(4,12,16,9)}
=
[0.0002,0.0016]\times[5{,}000{,}000,17{,}500{,}000]\times[0.926,0.937]\times[0.901,0.912].
\]
\\ \\
\textbf{Step 5: Assign weights for $(r,a)$}.
Scenario 10 assumes independence between $r$ and $a$, hence, the mass for each $(r,a)$ is
\[
w^{ra}_{ij}=0.05\times0.05=0.0025.
\]
\\ \\
\textbf{Step 6: Assign weights for $(p,q)$}.
Scenario~10 assumes Gaussian-copula dependence between \(p\) and \(q\) with \(\rho_{pq}=-0.85\). For the \((p,q)\) quantile cell \((i,j)\), the probability mass is
\[
w^{pq}_{ij}
=
C_{-0.85}\!\left(\frac{i}{20},\frac{j}{20}\right)
-C_{-0.85}\!\left(\frac{i-1}{20},\frac{j}{20}\right)
-C_{-0.85}\!\left(\frac{i}{20},\frac{j-1}{20}\right)
+C_{-0.85}\!\left(\frac{i-1}{20},\frac{j-1}{20}\right).
\]
This mass is assigned to the precise-CDF quantile cell, not to a p-box cell.
\\ \\
\textbf{Step 7: Construct the Fr\'echet-type cross-dependence set \(\mathcal Q_\times\).}

Use \(u=1,\ldots,400\) to index the \((r,a)\) block cells and \(v=1,\ldots,400\) to index the \((p,q)\) quantile cells. In Scenario~10, the within-block masses are already fixed before cross-dependence is introduced:
\[
w^{ra}_u=0.0025,\qquad u=1,\ldots,400,
\]
because \(r\) and \(a\) are independent, while
\[
w^{pq}_v,\qquad v=1,\ldots,400,
\]
is the non-uniform Gaussian-copula cell mass from Step~6. The unknown cross-dependence is represented by the \(400\times400\) matrix
\[
\Pi^\times=\{\Pi^\times_{uv}:u=1,\ldots,400,\;v=1,\ldots,400\},
\]
where \(\Pi^\times_{uv}\) is the joint mass assigned to \((r,a)\) cell \(u\) and \((p,q)\) cell \(v\). The admissible Fr\'echet-type cross-dependence set is
\[
\mathcal Q_\times
=
\left\{
\Pi^\times:
\Pi^\times_{uv}\ge0,\;
\sum_{v=1}^{400}\Pi^\times_{uv}=w^{ra}_u\ \forall u,\;
\sum_{u=1}^{400}\Pi^\times_{uv}=w^{pq}_v\ \forall v
\right\}.
\]
Thus \(\mathcal Q_\times\) is a probability-mass coupling set. Its row margins are the \((r,a)\) masses, and its column margins are the Gaussian-copula \((p,q)\) masses. Because the \((p,q)\) masses are not uniform, the column constraints are not all equal to \(0.0025\). They are exactly the values computed from the Gaussian copula in Step~6. This step is the dependence-uncertainty optimization; it should be kept distinct from the model-output optimization in Step~8.
\\ \\
\textbf{Step 8: Evaluate the model}
For Scenario~10, write \(\mathcal H^{\times}_{u,v}\) for the full parameter-space cell associated with block-cell pair \((u,v)\); this is the block-indexed version of the joint cell \(\mathcal J_{\mathbf{k},\mathbf{r}}\) in Section~\ref{sec:cross_frechet}, after the p-box and precise-CDF multi-indices are flattened into \(u\) and \(v\). 
The index \(u\) identifies one of the \(400\) \((r,a)\) cells, and the index \(v\) identifies one of the \(400\) \((p,q)\) quantile cells. 
The lower and upper endpoints of the \((r,a)\) cell \(u\) are written as
\[
[r_u^-,r_u^+]\quad\text{and}\quad [a_u^-,a_u^+],
\]
and the lower and upper endpoints of the \((p,q)\) quantile cell \(v\) as
\[
[p_v^-,p_v^+]\quad\text{and}\quad [q_v^-,q_v^+].
\]
The superscript ``\(-\)'' marks the lower endpoint of the interval, and the superscript ``\(+\)'' marks the upper endpoint.
The full hyperrectangle is
\[
\mathcal H^{\times}_{u,v}
=
[r_u^-,r_u^+]\times[a_u^-,a_u^+]\times[p_v^-,p_v^+]\times[q_v^-,q_v^+].
\]
If \(u\) corresponds to the first \((r,a)\) slice pair and \(v\) corresponds to the first \((p,q)\) slice pair, then
\[
\mathcal H^{\times}_{u,v}
=
[0.0002,0.0016]\times[5.000,10.263]\times[0.000,0.659]\times[0.000,0.734],
\]
where \(a\) is expressed in millions in the displayed interval table.
For every \(\mathcal H^{\times}_{u,v}\), the loss function is evaluated at all \(2^4=16\) corner combinations:
\[
(r,a,p,q)\in
\{r_u^-,r_u^+\}\times
\{a_u^-,a_u^+\}\times
\{p_v^-,p_v^+\}\times
\{q_v^-,q_v^+\}.
\]
Then define the cellwise lower and upper outputs as
\[
\underline Y_{uv}
=
\min_{(r,a,p,q)\in\mathrm{corners}(\mathcal H^{\times}_{u,v})}
L(10\mid r,a,p,q),
\]
and
\[
\overline Y_{uv}
=
\max_{(r,a,p,q)\in\mathrm{corners}(\mathcal H^{\times}_{u,v})}
L(10\mid r,a,p,q).
\]
These two values form the output interval induced by the full hyperrectangle \(\mathcal H^{\times}_{u,v}\). 
Step~9 uses these precomputed \(\underline Y_{uv}\) and \(\overline Y_{uv}\) values to construct the output CDF bounds.
\\ \\
\textbf{Step 9: Construct the output CDF bounds by solving threshold-specific linear programs.}
For a fixed threshold \(y\), the event \(Y\le y\) is guaranteed for cell \((u,v)\) if the upper cell output does not exceed the threshold:
\[
\overline Y_{uv}\le y.
\]
Similarly, the event \(Y\le y\) is possible for cell \((u,v)\) if the lower cell output does not exceed the threshold:
\[
\underline Y_{uv}\le y.
\]
Therefore define the two binary indicators
\[
I^-_{uv}(y)=\mathbf 1\{\overline Y_{uv}\le y\},
\qquad
I^+_{uv}(y)=\mathbf 1\{\underline Y_{uv}\le y\}.
\]
The lower output CDF is obtained by minimizing the guaranteed mass below \(y\) over the admissible cross-dependence set:
\[
\underline F_Y(y)
=
\min_{\Pi^\times\in\mathcal Q_\times}
\sum_{u=1}^{400}\sum_{v=1}^{400}
\Pi^\times_{uv} I^-_{uv}(y).
\]
Equivalently, this is the linear program
\[
\begin{aligned}
\text{minimize}_{\Pi^\times}\quad&
\sum_{u=1}^{400}\sum_{v=1}^{400}
\Pi^\times_{uv}\mathbf 1\{\overline Y_{uv}\le y\}\\
\text{subject to}\quad&
\Pi^\times_{uv}\ge0,\qquad u,v=1,\ldots,400,\\
&
\sum_{v=1}^{400}\Pi^\times_{uv}=w^{ra}_u,\qquad u=1,\ldots,400,\\
&
\sum_{u=1}^{400}\Pi^\times_{uv}=w^{pq}_v,\qquad v=1,\ldots,400.
\end{aligned}
\]
The upper output CDF is obtained by maximizing the possible mass below \(y\):
\[
\overline F_Y(y)
=
\max_{\Pi^\times\in\mathcal Q_\times}
\sum_{u=1}^{400}\sum_{v=1}^{400}
\Pi^\times_{uv} I^+_{uv}(y),
\]
or explicitly,
\[
\begin{aligned}
\text{maximize}_{\Pi^\times}\quad&
\sum_{u=1}^{400}\sum_{v=1}^{400}
\Pi^\times_{uv}\mathbf 1\{\underline Y_{uv}\le y\}\\
\text{subject to}\quad&
\Pi^\times_{uv}\ge0,\qquad u,v=1,\ldots,400,\\
&
\sum_{v=1}^{400}\Pi^\times_{uv}=w^{ra}_u,\qquad u=1,\ldots,400,\\
&
\sum_{u=1}^{400}\Pi^\times_{uv}=w^{pq}_v,\qquad v=1,\ldots,400.
\end{aligned}
\]
These two linear programs are solved at each grid value \(y\in\mathcal Y\). The resulting values form the lower and upper output CDF envelope for Scenario~10. For the exceedance threshold \(T=3{,}000{,}000\), the probability interval is computed as
\[
\underline P(Y>T)=1-\overline F_Y(T),
\qquad
\overline P(Y>T)=1-\underline F_Y(T).
\]
The median interval is obtained from the same CDF envelope: the lower median is the first grid value where \(\overline F_Y(y)\ge0.5\), and the upper median is the first grid value where \(\underline F_Y(y)\ge0.5\). In this way, Scenario~10 has two explicit optimization layers: bounded model optimization within each full cell to obtain \((\underline Y_{uv},\overline Y_{uv})\), followed by linear optimization over \(\Pi^\times\in\mathcal Q_\times\) to obtain the output CDF bounds.
\subsection{Comparative analysis}
\label{sec:compar}
\begin{figure}[!htbp]
\centering
\includegraphics[width=0.94\textwidth]{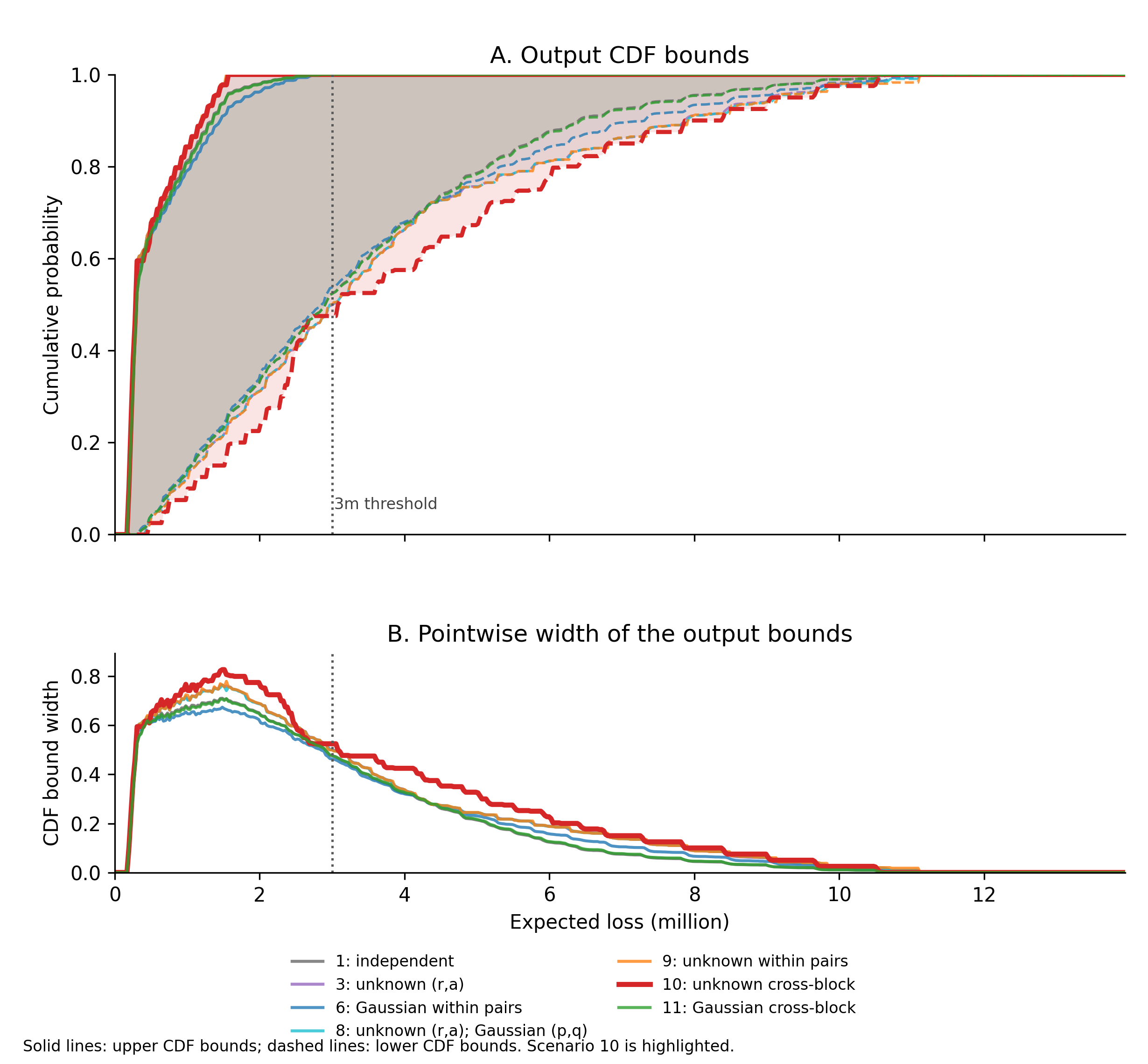}
\caption{Comparison of output uncertainty and pointwise bound width under selected dependence assumptions. Panel A shows the lower and upper PBA cumulative distribution bounds for Scenarios~1, 3, 6, 8, 9, 10, and~11 on a common expected-loss scale; solid curves denote upper CDF bounds and dashed curves denote lower CDF bounds. Panel B shows the corresponding pointwise CDF bound width, defined as the difference between the upper and lower CDF bounds at each expected-loss threshold. Scenario~10, which leaves cross-dependence between parameter blocks unspecified, has the largest maximum, mean, and integrated bound width across all eleven scenarios, although it is not pointwise widest at every threshold. This distinction is expected because CDF-bound width is a local threshold-specific quantity, whereas the overall uncertainty envelope is summarized by maximum, average, integrated, median, and exceedance-probability diagnostics.}
\label{fig:dependence_assumption_comparison}
\end{figure}

Table~\ref{tab:eleven_scenario_combined} lists the eleven scenarios and their corresponding result summaries.
The dependence assumptions primarily affect the pessimistic side (higher loss) of the uncertainty distribution rather than the optimistic side (lower loss). Across all eleven scenarios, the lower PBA median remains approximately \(0.308\) million because the optimistic admissible output distribution is governed mainly by the common lower-tail marginal supports of the input parameters. In contrast, the upper PBA median and the exceedance-probability interval \(P(Y>3\text{m})\) vary substantially across scenarios, indicating that dependence assumptions mainly alter the adverse tail behavior of the system.
The comparison between independence, Gaussian copulas, and Fr\'echet-type unknown dependence shows how stronger structural assumptions constrain the admissible uncertainty set. 
Gaussian-copula scenarios impose a specified dependence geometry and therefore generally produce narrower output envelopes. In contrast, Fr\'echet-type scenarios permit a much larger admissible set, allowing the optimization to place probability mass on more adverse configurations. 
Consequently, the Fr\'echet-type scenarios produce wider upper-tail uncertainty and larger exceedance-probability intervals than the corresponding Gaussian-copula scenarios. 
This difference is particularly more pronounced in Scenarios~8--10, where either the \((r,a)\) block, the cross-block structure, or both are modeled using Fr\'echet-type uncertainty.
Figure~\ref{fig:dependence_assumption_comparison} provides a graphical summary of selected scenarios on a common output scale and adds a direct diagnostic of the bound width. 
\begin{table}[!htbp]
\centering
\caption{Scenario specifications for the 11 scenarios under different dependence assumptions. PBA denotes probability bounds analysis; PSA denotes probabilistic sensitivity analysis; \((r,a)\) denotes the prevalence--outbreak-cost block; \((p,q)\) denotes the diagnostic sensitivity--specificity block; independent denotes product dependence; gaussian denotes Gaussian-copula dependence; frechet denotes Fr\'echet-type unknown dependence; and cross denotes dependence between the \((r,a)\) and \((p,q)\) blocks. PSA median is the median of the precise PSA output distribution. PBA median interval is the interval formed by the lower and upper admissible medians of the PBA output p-box. \(P(Y>3\text{m})\) denotes the exceedance-probability interval for expected loss \(Y\) exceeding \(3{,}000{,}000\). All median expected-loss values are reported in millions.}
\label{tab:eleven_scenario_combined}
\tiny
\setlength{\tabcolsep}{2.5pt}
\begin{adjustbox}{max width=\textwidth}
\begin{tabular}{rlllllccc}
\toprule
Scenario & PBA \((r,a)\) & PSA \((r,a)\) & PBA \((p,q)\) & PSA \((p,q)\) & Cross & PSA median & PBA median interval & \(P(Y>3\text{m})\) interval\\
\midrule
1 & independent & independent & independent & independent & none & 1.434 & (0.308, 2.912) & (0.000, 0.477)\\
2 & gaussian & gaussian & independent & independent & none & 1.454 & (0.308, 2.856) & (0.000, 0.464)\\
3 & frechet & gaussian & independent & independent & none & 1.425 & (0.308, 3.024) & (0.000, 0.502)\\
4 & independent & independent & gaussian & gaussian & none & 1.393 & (0.308, 2.912) & (0.000, 0.475)\\
5 & independent & independent & frechet & gaussian & none & 1.409 & (0.308, 2.940) & (0.000, 0.475)\\
6 & gaussian & gaussian & gaussian & gaussian & none & 1.410 & (0.308, 2.884) & (0.000, 0.463)\\
7 & gaussian & gaussian & frechet & gaussian & none & 1.426 & (0.308, 2.884) & (0.000, 0.462)\\
8 & frechet & gaussian & gaussian & gaussian & none & 1.378 & (0.308, 3.024) & (0.000, 0.500)\\
9 & frechet & gaussian & frechet & gaussian & none & 1.432 & (0.308, 2.968) & (0.000, 0.500)\\
10 & independent & independent & gaussian & gaussian & frechet & 1.427 & (0.308, 3.080) & (0.000, 0.525)\\
11 & gaussian & gaussian & gaussian & gaussian & gaussian & 1.418 & (0.308, 2.912) & (0.000, 0.477)\\
\bottomrule
\end{tabular}
\end{adjustbox}
\end{table}

\section{Discussion}
\subsection{Consequences of assuming independence without justification}
A central implication of the proposed framework is that independence should be treated as a substantive modeling assumption rather than as a convenient default. 
In both PSA and PBA, assuming independence determines how uncertainty is distributed across the joint parameter space. When this assumption is unjustified, the resulting uncertainty analysis may place probability mass on implausible combinations of parameters, exclude plausible combinations that would arise under dependence, or underestimate uncertainty in nonlinear functions of the parameters.
For example, if two beneficial treatment effects are positively associated, assuming independence may underrepresent the probability of jointly favorable or jointly unfavorable outcomes. Conversely, if two parameters are negatively associated because of a structural trade-off, independence may exaggerate the probability of extreme joint combinations. 
In both cases, the resulting uncertainty in the decision-relevant outcome may give a misleading impression of decision uncertainty.
The broader sensitivity-analysis literature has repeatedly emphasized that dependence among inputs changes both uncertainty propagation and attribution of output variability. 
Oakley and O'Hagan emphasize that uncertainty in model outputs is induced by the joint uncertainty distribution of model inputs, not only by their marginal distributions \citep{oakley2004}.
In applied risk and decision modeling, independence assumptions are common because model parameters are often estimated separately, elicited separately, or obtained from different data sources.
Neine et al. further showed in cost-effectiveness analysis that parameter correlation may affect uncertainty results and proposed methods for generating correlated non-normal input parameters \citep{neine2020}.
The two approaches developed in this article address the problem from different viewpoints. 
Copula-based PBA is appropriate when we can specify or estimate a plausible dependence structure. 
Fr\'echet bounds are appropriate when we cannot justify any particular dependence model and wish to avoid hidden assumptions. In both cases, the key principle is the same: independence should be used only when it is defensible, and sensitivity to dependence should be examined whenever dependence could materially affect model conclusions.
\subsection{Computational complexity after dropping the independence assumption}
A major implication of relaxing the independence assumption is the substantial increase in computational complexity. The computational burden is not only determined by the number of p-box parameters and the number of slices per p-box parameter, but also by the number of Monte Carlo samples for precise-CDF parameters, the cost of evaluating the black-box model, and the complexity of the optimization problems.
The first source of computational growth is the number of p-box-derived hyperrectangles. Let \(d_b\) denote the number of p-box parameters. If p-box parameter \(i\) is discretized into \(n_i\) slices, then the number of p-box-derived hyperrectangles under exact full-factorial propagation is
\[
|\mathcal{K}|=\prod_{i=1}^{d_b} n_i.
\]
This computational burden rapidly becomes significant even in moderate dimensions. For example, if there are 20 p-box parameters and each parameter is discretized into five slices, the number of hyperrectangles is \(5^{20}\approx 9.5\times10^{13}\).
Another source of computational growth is dependence mass assignment. 
Under copula-based dependence, the probability mass assigned to a single probability-space cell requires evaluation of the inclusion--exclusion formula over all \(2^{d_b}\) corners of that cell:
\[
P(\mathcal U_{\mathbf{k}})
=
\sum_{\mathbf{s}\in\{0,1\}^{d_b}}
(-1)^{d_b-\sum_i s_i}
C(u_1^{s_1},\ldots,u_{d_b}^{s_{d_b}}).
\]
This computing cost grows exponentially with the number of p-box parameters. 
The Fr\'echet-style formulation is also computationally demanding, albeit in a different way. 
Here, the cell mass \(p_{\mathbf{k}}\) becomes an optimization variable, 
and the output p-box is obtained through linear optimization over the admissible set \(\mathcal{Q}\). 
Thus, Fr\'echet propagation involves optimization over admissible joint probability-mass couplings \textit{in addition to} optimization of the black-box model over hyperrectangles. 
Parameters with precise CDFs contribute to computational complexity in a different way as well.
If \(N\) Monte Carlo draws are used for the precise-CDF parameters, then each p-box hyperrectangle must usually be propagated conditional on each draw \(\boldsymbol{\eta}^{(l)}\). 
The number of bounded model optimizations is therefore approximately
\[
2N\prod_{i=1}^{d_b}n_i,
\]
where the factor 2 corresponds to the minimization and maximization needed to obtain \(\underline{Y}_{\mathbf{k},l}\) and \(\overline{Y}_{\mathbf{k},l}\). 
Thus, even though precise-CDF parameters do not increase the number of p-box-derived hyperrectangles, they multiply the number of hyperrectangle optimizations by the number of samples used to integrate over their distributions. 
Fixed parameters do not add hyperrectangles and do not require sampling. Their computational contribution is indirect: they affect the numerical value and possibly the shape of the objective function being optimized.
One last contributor, and perhaps the dominant computational bottleneck, is the repeated black-box optimization step. For each hyperrectangle and each Monte Carlo draw of the precise-CDF parameters, the algorithm must solve
\[
\underline{Y}_{\mathbf{k},l}
=
\min_{\boldsymbol{\theta}_b\in\mathcal{H}_{\mathbf{k}}}
\mathcal{M}
(
\boldsymbol{\theta}_b,
\boldsymbol{\eta}^{(l)},
\boldsymbol{\phi}
),
\]
and
\[
\overline{Y}_{\mathbf{k},l}
=
\max_{\boldsymbol{\theta}_b\in\mathcal{H}_{\mathbf{k}}}
\mathcal{M}
(
\boldsymbol{\theta}_b,
\boldsymbol{\eta}^{(l)},
\boldsymbol{\phi}
).
\]
If the model is computationally expensive to evaluate, these repeated bounded optimizations will undoubtedly dominate runtime.
Hence, scalable computation approaches should be considered to significantly reduce the computational burden.
We sketch some ideas. 
First, instead of evaluating every combination of marginal slices (full-factorial design), the algorithm should focus on a sparse subset of representative hyperrectangles. 
This may change the effective computational scaling from exponential growth in the full Cartesian grid to approximately linear growth in the number of sampled hyperrectangles.
Second, to reduce the number of Monte Carlo samples we should consider the use of variance-reduced outer sampling, such as quasi-Monte Carlo Sobol draws, Latin hypercube sampling, antithetic sampling, or sparse quadrature when the number of precise-CDF parameters is small.
An adaptive approach can also be used: start with a modest number of precise-CDF draws, estimate the change in \(\underline{F}_Y\) and \(\overline{F}_Y\) as additional draws are added, and stop when the change is smaller than a prespecified tolerance. 
Third, low-influence precise-CDF parameters can also be screened through sensitivity analyses, and, when justified, fixed at representative values.
The propagation is naturally parallelizable because each hyperrectangle optimization is conditionally independent given the dependence model and Monte Carlo sample. 
Lastly, using meta-models or emulators in place of the black-box model may also reduce the need for repeated model evaluations, provided that emulator uncertainty is incorporated into the final output p-box rather than ignored \citep{ellis2020active}.
These aforementioned scalable approaches do not eliminate the intrinsic complexity of PBA under dependence uncertainty, but they substantially improve tractability and make high-dimensional dependence-sensitive uncertainty propagation feasible for practical applications.
\subsection{Decision-making}
One major implication of PBA-based uncertainty quantification is that the output of the analysis is no longer a single precise distribution of $Y$, but an output p-box characterized by lower and upper cumulative distribution bounds.
This changes the interpretation of decision-making fundamentally relative to conventional PSA. 
Under PSA, decision-making is grounded in the von Neumann--Morgenstern (VNM) expected utility framework, which assumes precise probability distributions for uncertain model parameters and therefore a unique induced distribution for the model outcome \citep{vonneumann1944theory,savage1954foundations}. 
In health economic evaluation, this usually leads to decision criteria based on expected net monetary benefit \citep{claxton1999probabilistic,fenwick2001representing}. 
PBA produces a family of admissible distributions,
\[
\mathcal F_Y
=
\left\{
F_Y:
\underline F_Y(y)
\le
F_Y(y)
\le
\overline F_Y(y)
\right\},
\]
so the expected utility itself becomes interval-valued,
\[
\underline E[Y]
\le
E[Y]
\le
\overline E[Y].
\]
Several alternative decision criteria arise naturally for interval-valued decision-relevant outcomes. 
One possibility is interval dominance, where strategy $A$ dominates strategy $B$ if
\[
\underline E(Y_A)
>
\overline E(Y_B),
\]
meaning that every admissible distribution for $A$ yields a larger expected outcome than every admissible distribution for $B$. 
Another possibility is maximality, which compares strategies over the entire admissible set of distributions and allows partially ordered preferences \citep{troffaes2014lower}. 
A more conservative alternative is robust or \(\Gamma\)-maximin decision-making over $d$ choices,
\[
\max_d
\min_{F\in\mathcal F_Y}
E_F[Y_d],
\]
which selects the strategy with the best worst-case expected performance under all admissible distributions \citep{gilboa1989maxmin}. 
In sum, any aforementioned decision-making framework can be readily integrated into PBA. 
Nevertheless, we should distinguish uncertainty quantification from the choice of normative decision models:  PBA characterizes the admissible uncertainty set implied by incomplete information and unresolved dependence assumptions, whereas the choice of decision criterion depends on the decision-makers' tolerance for ambiguity and conservatism. 
\subsection{Limitations}
This study has several limitations. First, as noted before, although the proposed framework generalizes PBA to include settings where accounting for dependence uncertainty matters, the computational burden may still be substantial for high-dimensional problems.
We offer some solutions in the above discussion.
However, a thorough investigation is needed to determine which scalable approaches are optimal.
Next, this study introduces a copula approach for modeling dependence under PBA. 
We do not provide guidance on how to select the appropriate copula model.
However, our characterization of copula and its application are general, hence, their utility is not restricted to a particular copula model.
Thirdly, although we introduce approaches that can accommodate a broad class of dependence assumptions, we do not provide practical guidance for selecting an appropriate dependence model in a given application. 
In many real-world settings, information about dependence is sparse, incomplete, or entirely unavailable. Consequently, results may remain sensitive to assumptions regarding the nature and strength of dependence.
Fourth, the illustrative application was intentionally designed to demonstrate the methodology rather than to represent the full complexity of real-world models. Although the example highlights the practical implications of dependence assumptions, additional applications involving more realistic models, multiple sources of evidence, and more complex dependence structures would strengthen the empirical evaluation of the framework.
Fifth, the framework assumes that the information used to construct uncertainty bounds is itself reliable and representative. If the underlying evidence base is biased (e.g., due to measurement error or not representative of the target population), the resulting uncertainty characterization may also be misleading regardless of conservativeness of the bounds. 
The framework quantifies uncertainty conditional on the available evidence but does not account for the robustness in the evidence.
Lastly, practical implementation still requires analyst judgment regarding uncertainty representation, discretization choices, optimization settings, and dependence assumptions. Although these choices are transparent and can be examined through sensitivity analyses, different modeling decisions may produce different uncertainty bounds, particularly when the available evidence is limited.

\subsection{Future directions}
Although the proposed framework broadens PBA to accommodate dependent uncertainty, several methodological and computational challenges remain.
First, practical guidance is needed for choosing among copula and Fr\'echet dependence models. In applied risk and decision modeling, dependence information is often partial: we may know the direction of the association, a plausible correlation range, or structural constraints, but not a full joint distribution. 
One future research topic is the development of decision rules or diagnostic workflows that map available dependence information to an appropriate dependence model. For example, pairwise correlation information may motivate a Gaussian or vine copula; qualitative tail-dependence information may motivate Archimedean or \(t\)-copulas; and absence of defensible dependence information may motivate Fr\'echet bounds.
Second, dependence elicitation remains underdeveloped. 
Existing elicitation methods in health economic modeling often focus on marginal distributions\citep{iskandar2021}.
Future work should develop elicitation protocols for dependence under sparse data, including elicitation of rank correlations, tail dependence, and feasible joint constraints. Such protocols should be designed for domain experts who may not be familiar with copula theory or imprecise probability.
Third, scalable algorithms are needed for high-dimensional dependent PBA. The curse of dimensionality affects both hyperrectangle enumeration and dependence-sensitive mass assignment. Sparse Sobol hyperrectangles, adaptive refinement, and surrogate modeling provide promising directions, but their statistical properties require further study. In particular, future research should quantify approximation error introduced by sparse hyperrectangle selection and develop stopping criteria indicating when additional hyperrectangles no longer materially change the output p-box.
Fourth, optimization over hyperrectangles remains a major bottleneck. Fast local optimizers may fail for multimodal black-box models, whereas global optimizers may be too slow for large propagation problems. Future implementations could combine several strategies: monotonicity screening, endpoint evaluation when justified, local optimization with multi-start initialization, surrogate-assisted optimization, and adaptive allocation of computational effort to influential hyperrectangles. Machine-learning emulators may also reduce repeated model evaluations, provided that emulator uncertainty is incorporated into the output p-box rather than ignored.
Fifth, Fr\'echet dependence propagation requires efficient optimization over admissible joint mass distributions. Although the discrete formulation leads to linear programming problems, the number of variables grows with the number of hyperrectangles. Future work should explore decomposition methods, optimal-transport formulations, constraint screening, and sparse-support approximations. These tools may make unknown-dependence PBA feasible for higher-dimensional models.

\section{Conclusion}
Uncertainty is an inevitable feature of models, and the available evidence is often insufficient to fully characterize all uncertain model parameters and their relationships. Conventional approaches frequently require assumptions that may not be supported by the underlying evidence, potentially leading to overconfident conclusions and an incomplete representation of uncertainty.
This study presents a framework that enables uncertainty analyses to more closely reflect the information that is actually available. By explicitly accommodating incomplete knowledge and avoiding reliance on unjustified assumptions, the proposed approach provides a more transparent characterization of the range of outcomes that remain plausible given current evidence.
\section*{Data Availability Statement}
The illustrative example uses numerical inputs described in the manuscript and supplementary material. Analysis code and derived data can be provided through the journal submission system upon request during review.

\bibliography{references}
\end{document}


\maketitle

\appendix

\section{Why the Fr\'echet coupling set preserves the marginals}\label{app:frechet-coupling-marginals}

\begin{theorem}[Fr\'echet admissible set under fixed marginals]
Let \(X\) and \(Z\) be finite-valued random elements with marginal probability masses
\[
P(X=u)=w_u^X,\qquad u=1,\ldots,m,
\]
and
\[
P(Z=v)=w_v^Z,\qquad v=1,\ldots,n,
\]
where \(w_u^X\ge0\), \(w_v^Z\ge0\), \(\sum_{u=1}^m w_u^X=1\), and \(\sum_{v=1}^n w_v^Z=1\). Define the coupling set
\[
\Gamma(w^X,w^Z)
=
\left\{
\Pi\in\mathbb R_+^{m\times n}:
\sum_{v=1}^{n}\Pi_{uv}=w_u^X,\;
\sum_{u=1}^{m}\Pi_{uv}=w_v^Z
\right\}.
\]
Then \(\Gamma(w^X,w^Z)\) is exactly the set of all joint distributions of \((X,Z)\) with marginals \(w^X\) and \(w^Z\).
\end{theorem}

\begin{proof}
If \(\Pi\) is a joint distribution of \((X,Z)\), then \(\Pi_{uv}=P(X=u,Z=v)\ge0\). Its row sums recover the marginal distribution of \(X\):
\[
\sum_{v=1}^{n}\Pi_{uv}
=
\sum_{v=1}^{n}P(X=u,Z=v)
=
P(X=u)
=
w_u^X.
\]
Similarly, its column sums recover the marginal distribution of \(Z\):
\[
\sum_{u=1}^{m}\Pi_{uv}
=
\sum_{u=1}^{m}P(X=u,Z=v)
=
P(Z=v)
=
w_v^Z.
\]
Therefore every valid joint distribution with the prescribed marginals belongs to \(\Gamma(w^X,w^Z)\). Conversely, if \(\Pi\in\Gamma(w^X,w^Z)\), then \(\Pi_{uv}\ge0\), and
\[
\sum_{u=1}^{m}\sum_{v=1}^{n}\Pi_{uv}
=
\sum_{u=1}^{m}w_u^X
=
1.
\]
Thus \(\Pi\) is a valid probability mass function on the product space. Its row and column sums are \(w^X\) and \(w^Z\), respectively, by construction. Hence \(\Gamma(w^X,w^Z)\) is precisely the set of all admissible couplings with the specified marginals.
\end{proof}

\begin{corollary}[Dependence uncertainty only]
Optimizing over \(\Gamma(w^X,w^Z)\) varies only the dependence structure between \(X\) and \(Z\). The marginal distributions remain fixed because the row and column constraints preserve \(w^X\) and \(w^Z\) exactly.
\end{corollary}

\section{Mathematical justification of the dependence-sensitive PBA construction}
\label{app:math-justification}

This appendix states the finite-discretized claims used by the dependence-sensitive PBA construction. The results apply to the probability-scale representation induced by the chosen slices, rather than to an uncountable limiting p-box problem. Sharpness is therefore understood with respect to the finite probability-mass coupling problem used by the algorithm \cite{iskandar2021pba,claxton2005psa,jackson2019voi}.

Using the probability-scale discretization defined in the main text, \(K\) is the number of p-box parameters, \(j=1,\ldots,n_i\) indexes probability slices for parameter \(i\), and
\[
m_i^j=d_i^j-c_i^j,\qquad
m_i^j\ge0,\qquad
\sum_{j=1}^{n_i}m_i^j=1.
\]
A joint slice selection is written \(\mathbf{k}=(k_1,\ldots,k_K)\in\mathcal K\), with \(k_i\in\{1,\ldots,n_i\}\). The corresponding probability-space cell and parameter-space hyperrectangle are
\[
\mathcal U_{\mathbf{k}}=\prod_{i=1}^K[c_i^{k_i},d_i^{k_i}],
\qquad
\mathcal H_{\mathbf{k}}=\prod_{i=1}^K[a_i^{k_i},b_i^{k_i}].
\]
As in the main text, fixed-weight cases use \(w_{\mathbf{k}}\) for the mass assigned to \(\mathcal U_{\mathbf{k}}\), whereas unknown-dependence cases use \(p_{\mathbf{k}}\) for a feasible cell-mass variable. The cross-dependence notation \(\mathcal Q_{bc}\) and \(p_{\mathbf{k},\mathbf{r}}\) extend the same convention to the joint p-box/precise-CDF cell index \((\mathbf{k},\mathbf{r})\). Copula-based masses use the probability-scale representation of copulas, whose one-dimensional marginals are uniform \cite{nelsen2006copulas}. Fr\'echet classes represent joint distributions or finite joint probability-mass couplings with fixed marginals \cite{ruschendorf1991frechet}. The p-box discretization itself can be interpreted as a finite random-set or focal-element approximation to the underlying p-box representation \cite{ferson2009pbox}.

\begin{proposition}[Preservation of marginal slice masses]
\label{prop:marginal-mass-preservation}
Assume that each marginal discretization is a finite partition of the probability scale. Then the independent product allocation, any valid copula-induced allocation on the same probability-scale partition, and any allocation in the Fr\'echet feasible set preserve the marginal slice masses:
\[
\sum_{\mathbf{k}:k_i=j} w_{\mathbf{k}}=m_i^j
\quad\text{or}\quad
\sum_{\mathbf{k}:k_i=j} p_{\mathbf{k}}=m_i^j
\]
for all \(i\) and \(j\), as appropriate.
\end{proposition}

\begin{proof}
Under independence,
\[
w_{\mathbf{k}}^I=\prod_{i=1}^{K}m_i^{k_i}.
\]
For a fixed parameter \(i\) and slice \(j\),
\[
\sum_{\mathbf{k}:k_i=j}w_{\mathbf{k}}^I
=
\sum_{k_{-i}}
m_i^j\prod_{r\ne i}m_r^{k_r}
=
m_i^j
\prod_{r\ne i}\sum_{k_r=1}^{n_r}m_r^{k_r}
=
m_i^j.
\]
For a copula-induced allocation, \(w_{\mathbf{k}}^C=P_C\{U\in\mathcal U_{\mathbf{k}}\}\). Summing over all cells with \(k_i=j\) unions all other coordinates over their full probability-scale partitions and leaves only \(U_i\in[c_i^j,d_i^j]\). Since the \(i\)-th copula marginal is uniform,
\[
\sum_{\mathbf{k}:k_i=j}w_{\mathbf{k}}^C
=
P_C\{U_i\in[c_i^j,d_i^j]\}
=
d_i^j-c_i^j
=
m_i^j.
\]
For the Fr\'echet formulation, this same equality is imposed directly as a constraint in \(\mathcal Q\). Hence all three constructions preserve the marginal slice masses.
\end{proof}

\begin{proposition}[Preservation of block-copula margins]
\label{prop:block-copula-margin-preservation}
Assume \(\alpha_u\ge0\), \(u=1,\ldots,U\), and \(\beta_v\ge0\), \(v=1,\ldots,V\), satisfy \(\sum_u\alpha_u=\sum_v\beta_v=1\).  Define cumulative masses
\[
A_u=\sum_{s=1}^{u}\alpha_s,
\qquad
B_v=\sum_{t=1}^{v}\beta_t,
\qquad
A_0=B_0=0.
\]
For any valid bivariate copula \(C_\times\), define
\[
\pi_{uv}
= C_\times(A_u,B_v)-C_\times(A_{u-1},B_v)-C_\times(A_u,B_{v-1})+C_\times(A_{u-1},B_{v-1}).
\]
Then \(\pi_{uv}\ge0\), \(\sum_v\pi_{uv}=\alpha_u\), \(\sum_u\pi_{uv}=\beta_v\), and \(\sum_{u,v}\pi_{uv}=1\).
\end{proposition}

\begin{proof}
Nonnegativity follows because \(\pi_{uv}\) is the probability assigned by the copula measure to the rectangle \((A_{u-1},A_u]\times(B_{v-1},B_v]\).  For a fixed \(u\), summing over \(v\) telescopes:
\[
\sum_{v=1}^{V}\pi_{uv}
=
C_\times(A_u,1)-C_\times(A_{u-1},1)-C_\times(A_u,0)+C_\times(A_{u-1},0).
\]
A copula has uniform margins and satisfies \(C_\times(a,1)=a\) and \(C_\times(a,0)=0\).  Hence
\[
\sum_{v=1}^{V}\pi_{uv}=A_u-A_{u-1}=\alpha_u.
\]
The proof of \(\sum_u\pi_{uv}=\beta_v\) is identical, using \(C_\times(1,b)=b\) and \(C_\times(0,b)=0\).  Summing either marginal identity over its index gives \(\sum_{u,v}\pi_{uv}=1\).
\end{proof}

\begin{proposition}[Non-emptiness of the Fr\'echet admissible set]
\label{prop:frechet-nonempty}
If \(m_i^j\ge0\) and \(\sum_{j=1}^{n_i}m_i^j=1\) for every p-box parameter \(i\), then the finite Fr\'echet feasible set
\[
\mathcal Q
=
\left\{
p:\;p_{\mathbf{k}}\ge0,\quad
\sum_{\mathbf{k}\in\mathcal K}p_{\mathbf{k}}=1,\quad
\sum_{\mathbf{k}:k_i=j}p_{\mathbf{k}}=m_i^j\ \text{for all }i,j
\right\}
\]
is non-empty.
\end{proposition}

\begin{proof}
Construct the product allocation
\[
p_{\mathbf{k}}=\prod_{i=1}^{K}m_i^{k_i}.
\]
It is nonnegative. Moreover,
\[
\sum_{\mathbf{k}\in\mathcal K}p_{\mathbf{k}}
=
\prod_{i=1}^{K}\sum_{j=1}^{n_i}m_i^j
=
1.
\]
For fixed \(i\) and \(j\),
\[
\sum_{\mathbf{k}:k_i=j}p_{\mathbf{k}}
=
m_i^j\prod_{r\ne i}\sum_{k_r=1}^{n_r}m_r^{k_r}
=
m_i^j.
\]
Thus \(p\in\mathcal Q\), proving non-emptiness.
\end{proof}

\begin{assumption}[Output enclosure over each hyperrectangle]
\label{ass:output-enclosure}
For every retained hyperrectangle \(\mathcal H_{\mathbf{k}}\) and, when present, every outer precise-CDF draw \(\boldsymbol{\eta}^{(l)}\), the computed values \(\underline Y_{\mathbf{k},l}\) and \(\overline Y_{\mathbf{k},l}\) satisfy
\[
\underline Y_{\mathbf{k},l}
\le
\mathcal M(\boldsymbol{\theta}_b,\boldsymbol{\eta}^{(l)},\boldsymbol{\phi})
\le
\overline Y_{\mathbf{k},l}
\]
for every admissible \(\boldsymbol{\theta}_b\in\mathcal H_{\mathbf{k}}\). If no precise-CDF parameters are sampled, the index \(l\) and the outer averaging below are omitted.
\end{assumption}

\begin{proposition}[Validity of the fixed-weight output p-box]
\label{prop:valid-output-pbox}
Assume Assumption~\ref{ass:output-enclosure}. Suppose \(w_{\mathbf{k}}\ge0\) and \(\sum_{\mathbf{k}\in\mathcal K}w_{\mathbf{k}}=1\). With \(N\) outer draws of the precise-CDF parameters, define
\[
\underline F_Y(y)
=
\frac1N\sum_{l=1}^{N}\sum_{\mathbf{k}\in\mathcal K}
w_{\mathbf{k}}\mathbf 1\{\overline Y_{\mathbf{k},l}\le y\},
\qquad
\overline F_Y(y)
=
\frac1N\sum_{l=1}^{N}\sum_{\mathbf{k}\in\mathcal K}
w_{\mathbf{k}}\mathbf 1\{\underline Y_{\mathbf{k},l}\le y\}.
\]
Then \((\underline F_Y,\overline F_Y)\) is a valid lower/upper CDF pair for the discretized output distribution induced by the weights \(w_{\mathbf{k}}\): both functions are nondecreasing, right-continuous step functions with values in \([0,1]\), and
\[
0\le \underline F_Y(y)\le \overline F_Y(y)\le1
\]
for every \(y\). The result applies to independence and specified copula cases by taking \(w_{\mathbf{k}}=w_{\mathbf{k}}^I\) or \(w_{\mathbf{k}}=w_{\mathbf{k}}^C\).
\end{proposition}

\begin{proof}
Each indicator function is nondecreasing in \(y\), and each weighted sum has nonnegative weights summing to one, so both functions are nondecreasing step functions bounded between zero and one. They are right-continuous because they are finite sums of right-continuous indicator functions of the form \(\mathbf 1\{a\le y\}\). Since \(\underline Y_{\mathbf{k},l}\le \overline Y_{\mathbf{k},l}\),
\[
\mathbf 1\{\overline Y_{\mathbf{k},l}\le y\}
\le
\mathbf 1\{\underline Y_{\mathbf{k},l}\le y\}
\]
for every \((\mathbf{k},l)\) and \(y\). Multiplying by \(w_{\mathbf{k}}/N\ge0\) and summing over \(\mathbf{k}\) and \(l\) gives
\[
\underline F_Y(y)\le \overline F_Y(y).
\]
Assumption~\ref{ass:output-enclosure} gives the lower and upper probability interpretation at each outer draw: if \(\overline Y_{\mathbf{k},l}\le y\), then all attainable outputs in cell \(\mathbf{k}\) at draw \(l\) are at most \(y\); if \(\underline Y_{\mathbf{k},l}>y\), then no attainable output in cell \(\mathbf{k}\) at draw \(l\) is at most \(y\). Therefore the two finite averages bound the CDF of any output distribution compatible with the within-cell enclosures.
\end{proof}

\begin{proposition}[Validity of the Fr\'echet output envelopes]
\label{prop:valid-frechet-output-envelope}
Assume Assumption~\ref{ass:output-enclosure} and suppose \(\mathcal Q\ne\varnothing\). Define
\[
\underline F_Y^{\,F}(y)
=
\min_{p\in\mathcal Q}
\frac1N\sum_{l=1}^{N}\sum_{\mathbf{k}\in\mathcal K}
p_{\mathbf{k}}\mathbf 1\{\overline Y_{\mathbf{k},l}\le y\},
\qquad
\overline F_Y^{\,F}(y)
=
\max_{p\in\mathcal Q}
\frac1N\sum_{l=1}^{N}\sum_{\mathbf{k}\in\mathcal K}
p_{\mathbf{k}}\mathbf 1\{\underline Y_{\mathbf{k},l}\le y\}.
\]
Then
\[
0\le \underline F_Y^{\,F}(y)\le \overline F_Y^{\,F}(y)\le1
\]
for every \(y\), and both functions are nondecreasing right-continuous step functions in \(y\).
\end{proposition}

\begin{proof}
For fixed \(p\in\mathcal Q\), both objectives are finite weighted sums of indicators with nonnegative weights summing to one; hence each lies in \([0,1]\). Taking a minimum or maximum over \(\mathcal Q\) preserves the outer bounds \(0\) and \(1\).

For every \(p\in\mathcal Q\),
\[
\frac1N\sum_{l=1}^{N}\sum_{\mathbf{k}} p_{\mathbf{k}}\mathbf 1\{\overline Y_{\mathbf{k},l}\le y\}
\le
\frac1N\sum_{l=1}^{N}\sum_{\mathbf{k}} p_{\mathbf{k}}\mathbf 1\{\underline Y_{\mathbf{k},l}\le y\}.
\]
The minimum of the left-hand objective over \(\mathcal Q\) cannot exceed the maximum of the right-hand objective over the same set, giving
\[
\underline F_Y^{\,F}(y)\le \overline F_Y^{\,F}(y).
\]
For monotonicity, let \(y_1\le y_2\). For every feasible \(p\), the lower-objective value at \(y_1\) is no greater than the lower-objective value at \(y_2\). Taking minima over the same feasible set yields
\[
\underline F_Y^{\,F}(y_1)\le \underline F_Y^{\,F}(y_2).
\]
The same argument with maxima gives
\[
\overline F_Y^{\,F}(y_1)\le \overline F_Y^{\,F}(y_2).
\]
Right-continuity follows because only finitely many indicator coefficient vectors occur as \(y\) varies, and each indicator is of the right-continuous form \(\mathbf 1\{a\le y\}\). Hence the pointwise minima and maxima over the same finite set of coefficient vectors are right-continuous step functions.
\end{proof}

\begin{theorem}[Nesting of copula-based bounds within Fr\'echet bounds]
\label{prop:copula-nested-in-frechet}
Suppose \(w^C=(w_{\mathbf{k}}^C:\mathbf{k}\in\mathcal K)\) are the cell masses induced by a valid copula on the same probability-scale partition \(\{\mathcal U_{\mathbf{k}}:\mathbf{k}\in\mathcal K\}\). Suppose the copula's one-dimensional marginals agree with the slice masses \(m_i^j\), and suppose \(\mathcal Q\) is the Fr\'echet feasible set determined by those same marginal slice masses. Then \(w^C\in\mathcal Q\). Consequently, for every threshold \(y\),
\[
\underline F_Y^{\,F}(y)
\le
\underline F_Y^{\,C}(y)
\le
\overline F_Y^{\,C}(y)
\le
\overline F_Y^{\,F}(y).
\]
\end{theorem}

\begin{proof}
By Proposition~\ref{prop:marginal-mass-preservation}, the copula-induced masses \(w_{\mathbf{k}}^C\) are nonnegative, sum to one, and satisfy
\[
\sum_{\mathbf{k}:k_i=j}w_{\mathbf{k}}^C=m_i^j
\]
for all \(i,j\). These are exactly the constraints defining \(\mathcal Q\), so \(w^C\in\mathcal Q\).

For the lower CDF, \(\underline F_Y^{\,F}(y)\) is the minimum of the same lower-bound objective over all \(p\in\mathcal Q\). The copula lower bound is that objective evaluated at the feasible point \(w^C\). Hence
\[
\underline F_Y^{\,F}(y)\le \underline F_Y^{\,C}(y).
\]
For the upper CDF, \(\overline F_Y^{\,F}(y)\) is the maximum of the same upper-bound objective over all \(p\in\mathcal Q\), and the copula upper bound is the objective evaluated at \(w^C\). Hence
\[
\overline F_Y^{\,C}(y)\le \overline F_Y^{\,F}(y).
\]
The middle inequality \(\underline F_Y^{\,C}(y)\le \overline F_Y^{\,C}(y)\) follows from Proposition~\ref{prop:valid-output-pbox}. Combining the three inequalities proves the result.
\end{proof}

\begin{theorem}[Threshold-wise sharpness of the Fr\'echet-style bounds]
\label{prop:frechet-sharpness}
Assume the finite discretized hyperrectangle representation, Assumption~\ref{ass:output-enclosure}, and \(\mathcal Q\ne\varnothing\). Suppose the only information imposed about the joint allocation of p-box slice masses is the collection of marginal constraints defining \(\mathcal Q\). Then, for each fixed threshold \(y\), the Fr\'echet-style lower and upper bounds are sharp over the finite admissible class \(\mathcal Q\). That is, no larger lower bound or smaller upper bound at that same \(y\) can be guaranteed without adding further dependence assumptions.
\end{theorem}

\begin{proof}
Fix \(y\). The set \(\mathcal Q\) is a closed and bounded polytope in a finite-dimensional Euclidean space, because it is defined by finitely many linear equalities and inequalities together with \(p_{\mathbf{k}}\ge0\) and \(\sum_{\mathbf{k}}p_{\mathbf{k}}=1\). Since \(\mathcal Q\ne\varnothing\), it is compact. The lower and upper objectives
\[
L(p;y)=\frac1N\sum_{l=1}^{N}\sum_{\mathbf{k}} p_{\mathbf{k}}\mathbf 1\{\overline Y_{\mathbf{k},l}\le y\},
\qquad
U(p;y)=\frac1N\sum_{l=1}^{N}\sum_{\mathbf{k}} p_{\mathbf{k}}\mathbf 1\{\underline Y_{\mathbf{k},l}\le y\}
\]
are linear in \(p\). Therefore, both extrema are attained.

Take \(p^\star\in\mathcal Q\) to minimize \(L(p;y)\). Any proposed lower bound \(\widetilde F_L(y)\) using only the same marginal slice information must hold for every feasible \(p\in\mathcal Q\). If
\[
\widetilde F_L(y)>\underline F_Y^{\,F}(y)=L(p^\star;y),
\]
then \(\widetilde F_L(y)\) fails for the feasible allocation \(p^\star\), which satisfies all available marginal information. Hence no larger lower bound is guaranteed.

Similarly, let \(p^\dagger\in\mathcal Q\) maximize \(U(p;y)\). If a proposed upper bound \(\widetilde F_U(y)\) based only on the same marginal information satisfies
\[
\widetilde F_U(y)<\overline F_Y^{\,F}(y)=U(p^\dagger;y),
\]
then it fails for the feasible allocation \(p^\dagger\). Hence no smaller upper bound is guaranteed. The result is threshold-wise because the optimizing allocation may depend on \(y\).
\end{proof}

\begin{remark}[Cross-dependence extension]
\label{rem:cross-dependence-extension}
The preceding results use \(p_{\mathbf{k}}\) and \(\mathcal Q\) to keep notation simple. The cross-dependence version replaces \(k\) by the joint index \((\mathbf{k},\mathbf{r})\), \(p_{\mathbf{k}}\) by \(p_{\mathbf{k},\mathbf{r}}\), and \(\mathcal Q\) by \(\mathcal Q_{bc}\). If additional elicited constraints are imposed, the same arguments apply after replacing \(\mathcal Q_{bc}\) by the constrained feasible set \(\mathcal Q_{bc,\Gamma}\), provided that this constrained set is non-empty. If it is empty, the elicited constraints are incompatible with the marginal discretization and must be revised.
\end{remark}

\begin{remark}[Existence of hyperrectangle extrema]
\label{rem:extrema-existence}
If each \(\mathcal H_{\mathbf{k}}\) is closed and bounded and \(\mathcal M\) is continuous on \(\mathcal H_{\mathbf{k}}\) for fixed \((\boldsymbol{\eta},\boldsymbol{\phi})\), then the extrema defining \(\underline Y_{\mathbf{k},l}\) and \(\overline Y_{\mathbf{k},l}\) exist by the Weierstrass extreme value theorem for each outer draw \(l\). If \(\mathcal M\) is discontinuous, stochastic, or simulation-based, then the optimization rule, random seed handling, and simulation precision must be specified as part of the numerical protocol.
\end{remark}

\begin{remark}[Discretization as approximation]
\label{rem:discretization-approximation}
The finite hyperrectangle construction is a discretized approximation to the underlying p-box representation. Increasing the number of slices refines the finite problem but increases computational cost. Therefore, mathematical sharpness in Theorem~\ref{prop:frechet-sharpness} refers to the finite discretized problem induced by the selected slices.
\end{remark}

\begin{remark}[Compatibility of specified dependence models]
\label{rem:copula-compatibility}
A specified dependence model must define a coherent joint distribution on the probability scale. Pairwise correlations or pairwise copulas are not sufficient by themselves unless they are embedded in a valid multivariate copula or vine construction. Otherwise, the proposed cell masses may fail to be nonnegative or may fail to preserve the intended marginals.
\end{remark}

\section{Bivariate inclusion--exclusion derivation for copula cell masses}
\label{app:bivariate-copula-derivation}

This appendix derives the bivariate version of the copula-based probability mass assigned to a probability-space rectangle. The result is used in the copula-based dependence construction in the main text to compute the mass of each discretized probability-space cell under a specified copula.
Consider two p-box parameters with probability-scale variables \(U_1\) and \(U_2\). Let
\[
\mathcal U_{j_1,j_2}
=
[c_1^{j_1},d_1^{j_1}]
\times
[c_2^{j_2},d_2^{j_2}]
\]
be one bivariate probability-space rectangle. If \(C\) is the copula of \((U_1,U_2)\), then
\[
C(u_1,u_2)
=
P(U_1\le u_1,\;U_2\le u_2).
\]
The mass assigned to \(\mathcal U_{j_1,j_2}\) is
\[
w_{j_1,j_2}^{C}
=
P(c_1^{j_1}<U_1\le d_1^{j_1},\;
  c_2^{j_2}<U_2\le d_2^{j_2}).
\]
This probability is obtained by starting with the lower-left cumulative probability at the upper-right corner, subtracting the two rectangles outside the target cell, and then adding back the lower-left corner that was subtracted twice. By inclusion--exclusion,
\[
\begin{aligned}
w_{j_1,j_2}^{C}
&=
C(d_1^{j_1},d_2^{j_2})
-
C(c_1^{j_1},d_2^{j_2})
-
C(d_1^{j_1},c_2^{j_2})
+
C(c_1^{j_1},c_2^{j_2}).
\end{aligned}
\]
Endpoint conventions do not affect the formula for continuous copulas. More generally, the expression should be read as the probability increment of the copula over the rectangle, so it remains the correct finite cell mass when the discretized probability-scale cells form a partition.

Thus, the bivariate copula mass is the alternating sum of the copula evaluated at the four rectangle corners. This expression is the two-dimensional case of the general \(K\)-dimensional inclusion--exclusion formula used to assign copula-induced masses to the probability-space hyperrectangles \(\mathcal U_{\mathbf{k}}\).

\section{P-box formulas from minimal data}
\label{app:pbox-formulas}
This appendix summarizes several free p-box constructions used when only partial information about the unknown CDF is available. To avoid conflict with the vector notation in the main text, write \(x\) for a generic scalar parameter value with support \(I=[a,b]\). The formulas below follow the minimal-data constructions described in \cite{iskandar2022early}; they make explicit how the lower- and upper-bounding functions depend on the available data \(\mathcal{D}\).
\subsection{Known minimum and maximum}
If \(\mathcal{D}=\{a,b\}\), then the lower-bounding function is
\begin{equation}
\underline{F}_{I}(x)=
\begin{cases}
0, & x < b,\\
1, & b \leq x,
\end{cases}
\label{eq:pbox-minmax-lbf}
\end{equation}
and the upper-bounding function is
\begin{equation}
\overline{F}_{I}(x)=
\begin{cases}
0, & x < a,\\
1, & a \leq x.
\end{cases}
\label{eq:pbox-minmax-ubf}
\end{equation}
\subsection{Known minimum, maximum, and median}
If \(\mathcal{D}=\{a,b,m\}\), where \(m\) is the median, then
\begin{equation}
\underline{F}_{I,m}(x)=
\begin{cases}
0, & x < m,\\
1/2, & m \leq x < b,\\
1, & b \leq x,
\end{cases}
\label{eq:pbox-minmaxmedian-lbf}
\end{equation}
and
\begin{equation}
\overline{F}_{I,m}(x)=
\begin{cases}
0, & x < a,\\
1/2, & a \leq x < m,\\
1, & m \leq x.
\end{cases}
\label{eq:pbox-minmaxmedian-ubf}
\end{equation}
\subsection{Known minimum, maximum, and mean}
If \(\mathcal{D}=\{a,b,\mu\}\), where \(\mu=E(X)\) for a generic scalar uncertain quantity \(X\) supported on \([a,b]\), then
\begin{equation}
\underline{F}_{I,\mu}(x)=
\begin{cases}
0, & x < \mu,\\
\dfrac{x-\mu}{x-a}, & \mu \leq x < b,\\
1, & b \leq x,
\end{cases}
\label{eq:pbox-minmaxmean-lbf}
\end{equation}
and
\begin{equation}
\overline{F}_{I,\mu}(x)=
\begin{cases}
0, & x < a,\\
\dfrac{b-\mu}{b-x}, & a \leq x < \mu,\\
1, & \mu \leq x.
\end{cases}
\label{eq:pbox-minmaxmean-ubf}
\end{equation}
\subsection{Known minimum, maximum, mean, and standard deviation}
The formula below assumes \(a<\mu<b\) and a feasible variance satisfying the bounded-support moment constraints. In particular, the supplied \(\sigma^2\) must be compatible with support \([a,b]\) and mean \(\mu\); otherwise the moment constraints are inconsistent and no probability distribution satisfies the stated inputs.
If \(\mathcal{D}=\{a,b,\mu,\sigma\}\), define
\begin{equation}
\xi_1=\mu-\frac{\sigma^2}{b-\mu},
\qquad
\xi_2=\mu+\frac{\sigma^2}{\mu-a}.
\label{eq:pbox-xi-definitions}
\end{equation}
The lower-bounding function is
\begin{equation}
\underline{F}_{I,\mu,\sigma}(x)=
\begin{cases}
0, & x < \xi_1,\\[4pt]
\dfrac{\sigma^2+(b-\mu)(x-\mu)}{(b-a)(x-a)}, & \xi_1 \leq x < \xi_2,\\[10pt]
\dfrac{(x-\mu)^2}{(x-\mu)^2+\sigma^2}, & \xi_2 \leq x < b,\\[10pt]
1, & b \leq x,
\end{cases}
\label{eq:pbox-minmaxmeansd-lbf}
\end{equation}
and the upper-bounding function is
\begin{equation}
\overline{F}_{I,\mu,\sigma}(x)=
\begin{cases}
0, & x < a,\\[4pt]
\dfrac{\sigma^2}{(x-\mu)^2+\sigma^2}, & a \leq x < \xi_1,\\[10pt]
\dfrac{(b-\mu)(b-a+\mu-x)-\sigma^2}{(b-a)(b-x)}, & \xi_1 \leq x < \xi_2,\\[10pt]
1, & \xi_2 \leq x.
\end{cases}
\label{eq:pbox-minmaxmeansd-ubf}
\end{equation}
\subsection{Combining multiple sources of partial information}
When several primitive p-boxes are available for the same parameter, the resulting p-box can be obtained by intersecting the constraints. If \(d=1,\ldots,D\) indexes the primitive p-boxes induced by different subsets of \(\mathcal{D}\), then the intersected lower- and upper-bounding functions are
\begin{equation}
\underline{F}(x)=\max_{d=1,\ldots,D}\underline{F}_d(x),
\label{eq:pbox-intersection-lbf}
\end{equation}
and
\begin{equation}
\overline{F}(x)=\min_{d=1,\ldots,D}\overline{F}_d(x),
\label{eq:pbox-intersection-ubf}
\end{equation}
respectively. This construction yields the tightest p-box implied by the intersection of the available partial-information constraints, provided the constraints are mutually compatible.

\section{Generalized PBA algorithm}
\label{app:generalized_algorithm}
The algorithm uses the model-output and parameter-partition notation introduced in the main text. All dependence specifications are stated on probability-scale variables: \(U_i\in[0,1]\) for p-box parameter \(\theta_i\), and \(V_h=F_{\eta_h}(\eta_h)\in[0,1]\) for precise-CDF parameter \(\eta_h\). For a probability-space cell \(A\), write \(\Delta_A C\) for the copula probability increment over \(A\), evaluated by inclusion--exclusion or numerical integration.
The algorithm below separates four objects that must not be conflated: probability-scale slices, parameter-space intervals, fixed cell weights from specified dependence, and feasible mass sets from unknown dependence.
\\ \\
Throughout the algorithm, \(I_i^j\) and \(J_h^s\) denote probability-scale slices; \(H_i^j\) denotes a slice-induced value interval derived from a p-box; \(G_h^s\) denotes a quantile cell derived from a precise CDF; \(\mathcal H_{\mathbf{k}}\) denotes a p-box-derived hyperrectangle; and \(\mathcal G_{\mathbf{r}}\) denotes a precise-CDF quantile-cell rectangle. These objects are kept distinct throughout the propagation algorithm.
\begin{enumerate}[leftmargin=*,itemsep=0.65em]

\item \textbf{Construct marginal p-boxes.}
For each p-box parameter \(\theta_i\), construct
\[
\mathcal D_i\longmapsto (\underline F_i,\overline F_i),
\qquad i=1,\ldots,K,
\]
with \(\underline F_i(x)\le \overline F_i(x)\) for all \(x\).  The generalized inverse convention used throughout the paper is
\[
F^{-1}(u)=\inf\{x:F(x)\ge u\}.
\]

\item \textbf{Slice each p-box on the probability scale.}
Choose \(n_i\) probability slices for \(\theta_i\).  For slice \(j=1,\ldots,n_i\), define
\[
I_i^j=[c_i^j,d_i^j],
\qquad
m_i^j=d_i^j-c_i^j,
\qquad
\sum_{j=1}^{n_i}m_i^j=1,
\]
and map the slice to the slice-induced p-box value interval
\[
H_i^j=[a_i^j,b_i^j]
=
[\overline F_i^{-1}(c_i^j),\underline F_i^{-1}(d_i^j)].
\]
Thus \(I_i^j\) is a probability-scale slice, \(H_i^j\) is the corresponding parameter-space interval, and \(m_i^j\) is the marginal slice mass.

\item \textbf{Construct p-box-derived hyperrectangles.}
For a p-box multi-index \(k=(k_1,\ldots,k_K)\in\mathcal K=\prod_{i=1}^K\{1,\ldots,n_i\}\), define
\[
\mathcal U_{\mathbf{k}}=\prod_{i=1}^K I_i^{k_i},
\qquad
\mathcal H_{\mathbf{k}}=\prod_{i=1}^K H_i^{k_i}.
\]
The exact finite algorithm uses all \(\mathbf{k}\in\mathcal K\).  A scalable approximation may use a retained subset \(\mathcal K_S\subseteq\mathcal K\) generated by a Sobol, random, Latin-hypercube, adaptive, or other space-filling design.  Such a sparse implementation is an approximation to the exact finite discretization and must report how retained cell weights are normalized or reweighted.

\item \textbf{Represent precise-CDF parameters.}
There are two implementation modes for each precise-CDF parameter \(\eta_h\).

\begin{enumerate}[leftmargin=1.5em,itemsep=0.35em]
\item \textbf{Outer-simulation mode.}
If \(\eta_h\) does not participate in a finite dependence construction, draw or integrate it using its precise CDF \(F_{\eta_h}\) in the usual outer-loop calculation. For draw \(l=1,\ldots,N\), write
\[
\eta_h^{(l)}=F_{\eta_h}^{-1}(V_h^{(l)}),
\qquad V_h^{(l)}\in[0,1].
\]

\item \textbf{Finite-cell mode.}
If \(\eta_h\) participates in a finite copula, Fr\'echet-type, or cross-dependence construction, discretize its probability scale. Choose \(n_h^\eta\) slices
\[
J_h^s=[e_h^s,f_h^s],
\qquad
\mu_h^s=f_h^s-e_h^s,
\qquad
s=1,\ldots,n_h^\eta,
\]
and map them to ordinary precise-CDF quantile cells
\[
G_h^s=[\alpha_h^s,\beta_h^s]
=
[F_{\eta_h}^{-1}(e_h^s),F_{\eta_h}^{-1}(f_h^s)].
\]
These \(G_h^s\) are not slice-induced p-box value intervals. They are quantile cells of the precise CDF \(F_{\eta_h}\), introduced only so that precise-CDF parameters can enter a finite dependence construction.

For a precise-CDF multi-index \(\mathbf{r}=(r_1,\ldots,r_M)\in\mathcal R=\prod_{h=1}^M\{1,\ldots,n_h^\eta\}\), define
\[
\mathcal V_{\mathbf{r}}=\prod_{h=1}^M J_h^{r_h},
\qquad
\mathcal G_{\mathbf{r}}=\prod_{h=1}^M G_h^{r_h}.
\]
\end{enumerate}

\item \textbf{Assign p-box cell masses under independence.}
If \(U_1,\ldots,U_K\) are mutually independent, assign
\[
w_{\mathbf{k}}=\prod_{i=1}^K m_i^{k_i},
\qquad \mathbf{k}\in\mathcal K.
\]

\item \textbf{Assign p-box cell masses under a specified p-box copula.}
If \(\boldsymbol U=(U_1,\ldots,U_K)\) has a specified copula \(C_b\), assign
\[
w_{\mathbf{k}}=C_b(\mathcal U_{\mathbf{k}})=\Delta_{\mathcal U_{\mathbf{k}}}C_b.
\]
The computed mass vector must satisfy \(w_{\mathbf{k}}\ge0\), \(\sum_{\mathbf{k}\in\mathcal K}w_{\mathbf{k}}=1\), and \(\sum_{\mathbf{k}:k_i=j}w_{\mathbf{k}}=m_i^j\), up to numerical tolerance.  If these checks fail, the specified dependence model or numerical integration is not compatible with the finite discretization.

\item \textbf{Assign masses under specified cross-dependence.}
If dependence is specified jointly for \((\boldsymbol U,\boldsymbol V)\), use one of the following finite representations.  All of them produce fixed weights; none involves a Fr\'echet optimization.

\begin{enumerate}[leftmargin=1.5em,itemsep=0.3em]
\item \textbf{Direct joint-copula representation.}
For full probability-space cell
\[
\mathcal A_{k,r}=\mathcal U_{\mathbf{k}}\times\mathcal V_r,
\]
assign
\[
\pi_{k,r}=C_{bc}(\mathcal A_{k,r})=\Delta_{\mathcal A_{k,r}}C_{bc}.
\]
This is the most direct finite representation when a coherent joint copula for all probability-scale variables is available.

\item \textbf{Conditional-copula representation.}
For outer draw \(\boldsymbol V^{(l)}=\boldsymbol v^{(l)}\), assign
\[
w_{\mathbf{k}\mid l}=P\{\boldsymbol U\in\mathcal U_{\mathbf{k}}\mid \boldsymbol V=\boldsymbol v^{(l)}\},
\qquad
\sum_{\mathbf{k}\in\mathcal K}w_{\mathbf{k}\mid l}=1.
\]
The output p-box is then averaged over the outer draws.  This representation is useful when conditional distributions of the p-box probability-scale variables are easier to compute than full joint cell probabilities.

\item \textbf{Block-copula representation.}
Suppose a first block has cells \(u=1,\ldots,U\) with fixed within-block masses \(\alpha_u\), and a second block has cells \(v=1,\ldots,V\) with fixed within-block masses \(\beta_v\).  Choose deterministic orderings of the block cells and define cumulative masses
\[
A_u=\sum_{s=1}^{u}\alpha_s,
\qquad
B_v=\sum_{t=1}^{v}\beta_t,
\qquad
A_0=B_0=0.
\]
For a specified cross-block copula \(C_\times\), define
\[
\pi_{uv}
= C_\times(A_u,B_v)-C_\times(A_{u-1},B_v)-C_\times(A_u,B_{v-1})+C_\times(A_{u-1},B_{v-1}).
\]
Then \(\sum_v\pi_{uv}=\alpha_u\), \(\sum_u\pi_{uv}=\beta_v\), and \(\sum_{u,v}\pi_{uv}=1\).  This representation is used in Scenario~11.  It deliberately applies the cross-block copula to cumulative within-block masses, not to uniformly spaced block-cell labels; otherwise the within-block dependence margins would be overwritten.
\end{enumerate}

\item \textbf{Represent unknown dependence by a Fr\'echet feasible set.}
If dependence among p-box parameters is unknown, replace fixed weights by the feasible mass set
\[
\mathcal Q_b=
\left\{
p:\;p_{\mathbf{k}}\ge0,
\sum_{\mathbf{k}\in\mathcal K} p_{\mathbf{k}}=1,
\sum_{\mathbf{k}:k_i=j}p_{\mathbf{k}}=m_i^j\ \text{for all }i,j
\right\}.
\]
If unknown cross-dependence is between p-box cells \(\mathbf{k}\) and precise-CDF cells \(\mathbf{r}\), use finite-cell mode for the precise-CDF parameters and define
\[
\mathcal Q_{bc}=
\left\{
p:\;p_{\mathbf{k},\mathbf{r}}\ge0,
\sum_{\mathbf{k}\in\mathcal K}\sum_{\mathbf{r}\in\mathcal R}p_{\mathbf{k},\mathbf{r}}=1,
\sum_{\substack{\mathbf{k}\in\mathcal K\\ k_i=j}}\sum_{\mathbf{r}\in\mathcal R} p_{\mathbf{k},\mathbf{r}}=m_i^j,
\sum_{\substack{\mathbf{r}\in\mathcal R\\ r_h=s}}\sum_{\mathbf{k}\in\mathcal K} p_{\mathbf{k},\mathbf{r}}=\mu_h^s
\right\}.
\]
Additional elicited information may be imposed only if it is mathematically compatible with the discretized margins.  Linear restrictions lead to a smaller polytope \(\mathcal Q_{bc,\Gamma}\).  Nonlinear restrictions, such as some rank-correlation constraints, may require nonlinear optimization and should not be described as a linear-programming step unless they have been converted to linear constraints.

\item \textbf{Optimize the black-box over each retained parameter-space cell.}
For outer-simulation mode, compute, for every retained p-box cell \(k\) and draw \(l\),
\[
\underline Y_{\mathbf{k},l}=\min_{\boldsymbol{\theta}_b\in\mathcal H_{\mathbf{k}}}\mathcal M(\boldsymbol{\theta}_b,\boldsymbol{\eta}^{(l)},\boldsymbol{\phi}),
\qquad
\overline Y_{\mathbf{k},l}=\max_{\boldsymbol{\theta}_b\in\mathcal H_{\mathbf{k}}}\mathcal M(\boldsymbol{\theta}_b,\boldsymbol{\eta}^{(l)},\boldsymbol{\phi}).
\]
For finite joint cell \((\mathbf{k},\mathbf{r})\), compute
\[
\underline Y_{\mathbf{k},\mathbf{r}}=\min_{\boldsymbol{\theta}_b\in\mathcal H_{\mathbf{k}},\;\boldsymbol{\eta}\in\mathcal G_{\mathbf{r}}}\mathcal M(\boldsymbol{\theta}_b,\boldsymbol{\eta},\boldsymbol{\phi}),
\qquad
\overline Y_{\mathbf{k},\mathbf{r}}=\max_{\boldsymbol{\theta}_b\in\mathcal H_{\mathbf{k}},\;\boldsymbol{\eta}\in\mathcal G_{\mathbf{r}}}\mathcal M(\boldsymbol{\theta}_b,\boldsymbol{\eta},\boldsymbol{\phi}).
\]
Endpoint or corner evaluation is valid only when monotonicity or another exact extremum argument is available for the cell.  Otherwise, the implementation must use a bounded optimizer and report optimizer settings and tolerances.

\item \textbf{Construct the output p-box.}
\label{algstep:output}
For fixed p-box weights with outer-simulation draws, including independence or a specified p-box copula, compute
\[
\underline F_Y(y)=\frac1N\sum_{l=1}^N\sum_{\mathbf{k}\in\mathcal K} w_{\mathbf{k}}\mathbf 1\{\overline Y_{\mathbf{k},l}\le y\},
\qquad
\overline F_Y(y)=\frac1N\sum_{l=1}^N\sum_{\mathbf{k}\in\mathcal K} w_{\mathbf{k}}\mathbf 1\{\underline Y_{\mathbf{k},l}\le y\}.
\]
If no precise-CDF parameters are sampled, omit the outer average over \(l\).
For conditional weights, replace \(w_{\mathbf{k}}\) by \(w_{\mathbf{k}\mid l}\):
\[
\underline F_Y(y)=\frac1N\sum_{l=1}^N\sum_{\mathbf{k}\in\mathcal K} w_{\mathbf{k}\mid l}\mathbf 1\{\overline Y_{\mathbf{k},l}\le y\},
\qquad
\overline F_Y(y)=\frac1N\sum_{l=1}^N\sum_{\mathbf{k}\in\mathcal K} w_{\mathbf{k}\mid l}\mathbf 1\{\underline Y_{\mathbf{k},l}\le y\}.
\]
For fixed finite joint weights \(\pi_c\), where \(c=(\mathbf{k},\mathbf{r})\) or \(c=(u,v)\) depending on the cell construction, compute
\[
\underline F_Y(y)=\sum_c\pi_c\mathbf 1\{\overline Y_c\le y\},
\qquad
\overline F_Y(y)=\sum_c\pi_c\mathbf 1\{\underline Y_c\le y\}.
\]
For unknown p-box dependence with outer-simulation draws over precise-CDF parameters, use the already-computed quantities \(\underline Y_{\mathbf{k},l}\) and \(\overline Y_{\mathbf{k},l}\) from the cellwise model-optimization step. The optimization variable is the admissible mass vector \(p=(p_{\mathbf{k}})_{\mathbf{k}\in\mathcal K}\in\mathcal Q_b\). For each threshold \(y\), form
\[
\underline F_Y(y;p)=\frac1N\sum_{l=1}^N\sum_{\mathbf{k}\in\mathcal K}p_{\mathbf{k}}\mathbf 1\{\overline Y_{\mathbf{k},l}\le y\},
\qquad
\overline F_Y(y;p)=\frac1N\sum_{l=1}^N\sum_{\mathbf{k}\in\mathcal K}p_{\mathbf{k}}\mathbf 1\{\underline Y_{\mathbf{k},l}\le y\}.
\]
Then solve the two linear programs
\[
\underline F_Y(y)=\min_{p\in\mathcal Q_b}\underline F_Y(y;p),
\qquad
\overline F_Y(y)=\max_{p\in\mathcal Q_b}\overline F_Y(y;p).
\]
For unknown dependence represented by a finite joint cell system, solve
\[
\underline F_Y(y)=\min_{q\in\mathcal Q}\sum_c q_c\mathbf 1\{\overline Y_c\le y\},
\qquad
\overline F_Y(y)=\max_{q\in\mathcal Q}\sum_c q_c\mathbf 1\{\underline Y_c\le y\},
\]
where \(c=(\mathbf{k},\mathbf{r})\) for joint p-box/precise-CDF cells, \(q_c\) is only a generic placeholder for the cell-mass variable in that finite system, and \(\mathcal Q\) is the relevant feasible set, such as \(\mathcal Q_{bc}\) or \(\mathcal Q_{bc,\Gamma}\).

\item \textbf{Return the output p-box.}
Return
\[
\mathcal P_Y=(\underline F_Y,\overline F_Y)
\]
together with the cell definitions, cell weights or feasible mass set, dependence specification, optimizer settings, tolerance checks, and any sparse-subset approximation rule.
\end{enumerate}

\section{Computational model}
\label{app:example}
Following the article, the numerical model uses
\[
c(m)=1000-2000m+1000m^2,
\]
\[
a(d)=
\begin{cases}
0, & d=0,\\
a, & d\ge 1,
\end{cases}
\]
where \(a\) is treated as an uncertain outbreak-cost input in the illustrative analysis.
and
\[
n=250,\qquad t(n)=400n=100{,}000.
\]

The diagnostic sensitivity \(p\) and specificity \(q\) are treated as model inputs; in the illustrative analysis they are assigned precise beta marginal distributions. Write \(T\) for the event that the herd is terminated and \(T^c\) for the event that the herd passes inspection. Conditional on \(d\) diseased animals in the herd, the loss is
\[
L(m,d,p,q,c,a,t)
=
c(m)+t(n)P(T\mid d)+a(d)P(T^c\mid d).
\]
Equivalently,
\[
L(m,d,p,q,c,a,t)
=
c(m)+t(n)+\{a(d)-t(n)\}P(T^c\mid d).
\]
If the \(m\) tested animals are sampled without replacement, the number \(z\) of diseased animals in the tested sample follows a hypergeometric distribution. Hence the probability that the herd passes inspection is
\[
P(T^c\mid d)
=
\sum_{z}
(1-p)^z q^{m-z}
\frac{\binom{d}{z}\binom{n-d}{m-z}}{\binom{n}{m}},
\]
where the summation is taken over feasible values of \(z\). If each animal has infection probability \(r\), the number of diseased animals is modeled as
\[
P(d\mid r)
=
\binom nd r^d(1-r)^{n-d}.
\]
The expected loss for testing policy \(m\) is therefore
\[
L(m\mid r)
=
\sum_{d=0}^{n}
L(m,d,p,q,c,a,t)P(d\mid r).
\]

\section{Scenario-specific implementation details with numerical values}
\label{app:eleven_scenario_algorithms}

This appendix gives the numerical objects needed to reproduce the illustrative analysis without repeating the same algorithmic steps for every scenario. Table~\ref{tab:stress_exact_slice_intervals} reports the finite probability-slice intervals used in the implementation. The \(r\)- and \(a\)-columns are p-box-induced parameter intervals, whereas the \(p\)- and \(q\)-columns are beta-quantile intervals for precise-CDF parameters. The main text lists the scenario-specific dependence assumptions and output summaries.

\begin{table}[!htbp]
\centering
\caption{Exact 20-slice interval table used in the stress-test example. The \(a\)-intervals are shown in millions. The \(r\)- and \(a\)-columns are p-box-induced intervals, whereas the \(p\)- and \(q\)-columns are precise-CDF beta quantile intervals used for finite dependence calculations.}
\label{tab:stress_exact_slice_intervals}
\tiny
\setlength{\tabcolsep}{2.5pt}
\begin{adjustbox}{max width=\textwidth}
\begin{tabular}{rccccc}
\toprule
Slice \(j\) & Probability interval & \(r\)-interval & \(a\)-interval & \(p\)-interval & \(q\)-interval\\
\midrule
1 & [0.00,0.05] & [0.0002,0.0016] & [5.000,10.263] & [0.000,0.659] & [0.000,0.734]\\
2 & [0.05,0.10] & [0.0002,0.0016] & [5.000,10.556] & [0.659,0.712] & [0.734,0.785]\\
3 & [0.10,0.15] & [0.0002,0.0016] & [5.000,10.882] & [0.712,0.747] & [0.785,0.816]\\
4 & [0.15,0.20] & [0.0002,0.0016] & [5.000,11.250] & [0.747,0.773] & [0.816,0.840]\\
5 & [0.20,0.25] & [0.0002,0.0016] & [5.000,11.667] & [0.773,0.794] & [0.840,0.859]\\
6 & [0.25,0.30] & [0.0002,0.0016] & [5.000,12.143] & [0.794,0.812] & [0.859,0.875]\\
7 & [0.30,0.35] & [0.0002,0.0016] & [5.000,12.692] & [0.812,0.829] & [0.875,0.888]\\
8 & [0.35,0.40] & [0.0002,0.0016] & [5.000,13.333] & [0.829,0.843] & [0.888,0.901]\\
9 & [0.40,0.45] & [0.0002,0.0016] & [5.000,14.091] & [0.843,0.857] & [0.901,0.912]\\
10 & [0.45,0.50] & [0.0002,0.0016] & [5.000,15.000] & [0.857,0.870] & [0.912,0.922]\\
11 & [0.50,0.55] & [0.0002,0.0032] & [5.000,16.111] & [0.870,0.882] & [0.922,0.931]\\
12 & [0.55,0.60] & [0.0016,0.0032] & [5.000,17.500] & [0.882,0.893] & [0.931,0.940]\\
13 & [0.60,0.65] & [0.0016,0.0032] & [5.000,19.286] & [0.893,0.904] & [0.940,0.948]\\
14 & [0.65,0.70] & [0.0016,0.0032] & [5.000,20.000] & [0.904,0.915] & [0.948,0.956]\\
15 & [0.70,0.75] & [0.0016,0.0032] & [5.714,20.000] & [0.915,0.926] & [0.956,0.963]\\
16 & [0.75,0.80] & [0.0016,0.0032] & [6.667,20.000] & [0.926,0.937] & [0.963,0.970]\\
17 & [0.80,0.85] & [0.0016,0.0032] & [7.500,20.000] & [0.937,0.948] & [0.970,0.977]\\
18 & [0.85,0.90] & [0.0016,0.0032] & [8.235,20.000] & [0.948,0.960] & [0.977,0.984]\\
19 & [0.90,0.95] & [0.0016,0.0032] & [8.889,20.000] & [0.960,0.974] & [0.984,0.991]\\
20 & [0.95,1.00] & [0.0016,0.0032] & [9.474,20.000] & [0.974,1.000] & [0.991,1.000]\\
\bottomrule
\end{tabular}
\end{adjustbox}
\end{table}

The same finite cell system is used across scenarios. Let \(i,j\in\{1,\ldots,20\}\) index the \((r,a)\) slices and \(k,\ell\in\{1,\ldots,20\}\) index the \((p,q)\) quantile slices. The within-block product mass is
\[
w^{\mathrm{prod}}_{ij}=0.05\times0.05=0.0025.
\]
When a Gaussian copula with parameter \(\rho\) is specified for a block, the corresponding \(20\times20\) cell mass is
\[
w^{(\rho)}_{ij}
=
C_{\rho}\!\left(\frac{i}{20},\frac{j}{20}\right)
-
C_{\rho}\!\left(\frac{i-1}{20},\frac{j}{20}\right)
-
C_{\rho}\!\left(\frac{i}{20},\frac{j-1}{20}\right)
+
C_{\rho}\!\left(\frac{i-1}{20},\frac{j-1}{20}\right).
\]
When within-block dependence is treated as unknown, the admissible block-mass matrix is
\[
\Pi^{\mathrm{block}}
=
\left\{
\pi_{ij}:
\pi_{ij}\ge0,\quad
\sum_{j=1}^{20}\pi_{ij}=0.05,\quad
\sum_{i=1}^{20}\pi_{ij}=0.05
\right\}.
\]
The lower and upper CDF bounds are then obtained by optimizing the appropriate threshold-specific linear objective over the specified feasible mass set.

For Scenario~10, the \((r,a)\) block masses are fixed at \(w_u^{ra}=0.0025\), while the \((p,q)\) block masses \(w_v^{pq}\) are obtained from the Gaussian copula with \(\rho_{pq}=-0.85\). Unknown cross-dependence between the two blocks is represented by
\[
\mathcal Q_{\times}
=
\left\{
\Pi^\times:
\Pi^\times_{uv}\ge0,\quad
\sum_{v=1}^{400}\Pi^\times_{uv}=w_u^{ra},\quad
\sum_{u=1}^{400}\Pi^\times_{uv}=w_v^{pq}
\right\}.
\]
For threshold \(y\), the Scenario~10 output bounds are
\[
\underline F_Y(y)
=
\min_{\Pi^\times\in\mathcal Q_\times}
\sum_{u=1}^{400}\sum_{v=1}^{400}
\Pi^\times_{uv}\mathbf 1\{\overline Y_{uv}\le y\},
\]
and
\[
\overline F_Y(y)
=
\max_{\Pi^\times\in\mathcal Q_\times}
\sum_{u=1}^{400}\sum_{v=1}^{400}
\Pi^\times_{uv}\mathbf 1\{\underline Y_{uv}\le y\}.
\]

For Scenario~11, the within-block masses \(w_u^{ra}\) and \(w_v^{pq}\) are both fixed by Gaussian copulas. A Gaussian cross-block copula is then applied to the cumulative block-cell masses rather than to uniformly spaced block labels. Specifically, define
\[
A_u=\sum_{s=1}^{u}w_s^{ra},
\qquad
B_v=\sum_{t=1}^{v}w_t^{pq},
\qquad
A_0=B_0=0.
\]
With \(C_\times=C_{\mathrm{Gauss}}(\rho_\times=0.6)\), the full-cell mass is
\[
\pi_{uv}
=
C_\times(A_u,B_v)
-
C_\times(A_{u-1},B_v)
-
C_\times(A_u,B_{v-1})
+
C_\times(A_{u-1},B_{v-1}).
\]
This construction preserves the intended within-block margins:
\[
\sum_{v=1}^{400}\pi_{uv}=w_u^{ra},
\qquad
\sum_{u=1}^{400}\pi_{uv}=w_v^{pq}.
\]

\begin{table}[!htbp]
\centering
\caption{Mapping from the generalized algorithm in Appendix~\ref{app:generalized_algorithm} to the numerical implementation of Scenario~10. The table defines the Scenario~10 symbols used in the worked example, including the p-box block \((r,a)\), the precise-CDF block \((p,q)\), the within-block masses \(w^{ra}\) and \(w^{pq}\), and the cross-block feasible set \(\mathcal Q_\times\).}
\label{tab:scenario10_general_algorithm_mapping}
\scriptsize
\begin{tabular}{lll}
\toprule
Generalized object & Scenario~10 instantiation & Numerical value or construction\\
\midrule
Model output & \(Y\) & \(L(10\mid r,a,p,q)\)\\
P-box parameters & \(\boldsymbol{\theta}_b\) & \((r,a)\) after 20-slice p-box discretization\\
Precise-CDF parameters & \(\boldsymbol{\eta}\) & \((p,q)\) after 20-slice beta-quantile discretization\\
Decision & \(m\) & \(m=10\) for propagation; \(m=0,\ldots,20\) for decision analysis\\
\(r\) information & \(\mathcal D_r\) & \(\{0.0002,0.0016,0.0032\}\)\\
\(a\) information & \(\mathcal D_a\) & \(\{5{,}000{,}000,10{,}000{,}000,20{,}000{,}000\}\)\\
Diagnostic sensitivity & \(p\) & \(\mathrm{Beta}(10.2,1.8)\), \(E(p)=0.85\)\\
Diagnostic specificity & \(q\) & \(\mathrm{Beta}(10.8,1.2)\), \(E(q)=0.90\)\\
Within-pair dependence & \(C_{ra}\) & Independence\\
Within-pair dependence & \(C_{pq}\) & Gaussian copula, \(\rho_{pq}=-0.85\)\\
Cross-pair dependence & \(\mathcal Q_\times\) & Fr\'echet-type unknown cross-dependence over \(\Pi^\times_{uv}\)\\
Full cell system & \(\mathcal H_{\mathbf{k}}\times\mathcal G_{\mathbf{r}}\) & \(400\times400=160{,}000\) full cells\\
\bottomrule
\end{tabular}
\end{table}

The mapping in Table~\ref{tab:scenario10_general_algorithm_mapping} records how Scenario~10 instantiates the generalized algorithm. It is included to make the selected scenario reproducible without repeating the entire eleven-scenario step list.

\bibliography{references}